\documentclass[sigconf]{acmart}

\graphicspath{{figures/}}
\DeclareMathOperator*{\argmax}{arg\,max}
\usepackage{balance}
\usepackage{booktabs} 
\usepackage{subcaption}

\setcopyright{none} 

\acmDOI{}
\acmISBN{}

\acmConference[Anonymous Submission to ACM CCS 2019]{ACM Conference on Computer and Communications Security}{November 2019}{London, UK}
\acmYear{2019}
\copyrightyear{2019}

\renewcommand\footnotetextcopyrightpermission[1]{} 
\begin{document}
\title{Procedural Noise Adversarial Examples for Black-Box Attacks on Deep Convolutional Networks}
\renewcommand{\shorttitle}{Procedural Attack on Deep Learning}

\author{Kenneth T. Co}
\affiliation{
	\institution{Imperial College London}
}
\email{k.co@imperial.ac.uk}

\author{Luis Mu\~noz-Gonz\'alez}
\affiliation{
	\institution{Imperial College London}
}
\email{l.munoz@imperial.ac.uk}

\author{Sixte de Maupeou}
\affiliation{
	\institution{Imperial College London}
}
\email{sd4215@imperial.ac.uk}

\author{Emil C. Lupu}
\affiliation{
	\institution{Imperial College London}
}
\email{e.c.lupu@imperial.ac.uk}

\renewcommand{\shortauthors}{Co et al.}

\begin{abstract}
Deep Convolutional Networks (DCNs) have been shown to be vulnerable to adversarial examples---perturbed inputs specifically designed to produce intentional errors in the learning algorithms at test time. Existing input-agnostic adversarial perturbations exhibit interesting visual patterns that are currently unexplained. In this paper, we introduce a structured approach for generating Universal Adversarial Perturbations (UAPs) with procedural noise functions. Our approach unveils the systemic vulnerability of popular DCN models like Inception v3 and YOLO v3, with single noise patterns able to fool a model on up to 90\% of the dataset. Procedural noise allows us to generate a distribution of UAPs with high universal evasion rates using only a few parameters. Additionally, we propose Bayesian optimization to efficiently learn procedural noise parameters to construct inexpensive untargeted black-box attacks. We demonstrate that it can achieve an average of less than 10 queries per successful attack, a 100-fold improvement on existing methods. We further motivate the use of input-agnostic defences to increase the stability of models to adversarial perturbations. The universality of our attacks suggests that DCN models may be sensitive to aggregations of low-level class-agnostic features. These findings give insight on the nature of some universal adversarial perturbations and how they could be generated in other applications.
\end{abstract}



\maketitle

\section{Introduction}
Advances in computation and machine learning have enabled deep learning methods to become the favoured algorithms for various tasks such as computer vision \cite{krizhevsky2012imagenet}, malware detection \cite{saxe2015deep}, and speech recognition \cite{hinton2012deep}. Deep Convolutional Networks (DCN) achieve human-like or better performance in some of these applications. Given their increased use in safety-critical and security applications such as autonomous vehicles \cite{bojarski2016end, wu2017squeezedet, tian2018deeptest}, intrusion detection \cite{javaid2016deep, kang2016intrusion}, malicious string detection \cite{saxe2017expose}, and facial recognition \cite{sun2013deep, lopes2017facial}, it is important to ensure that such algorithms are robust to malicious adversaries. Yet despite the prevalence of neural networks, their vulnerabilities are not yet fully understood.

\begin{figure}[htb]
	\centering
	\includegraphics[clip, width = \columnwidth]{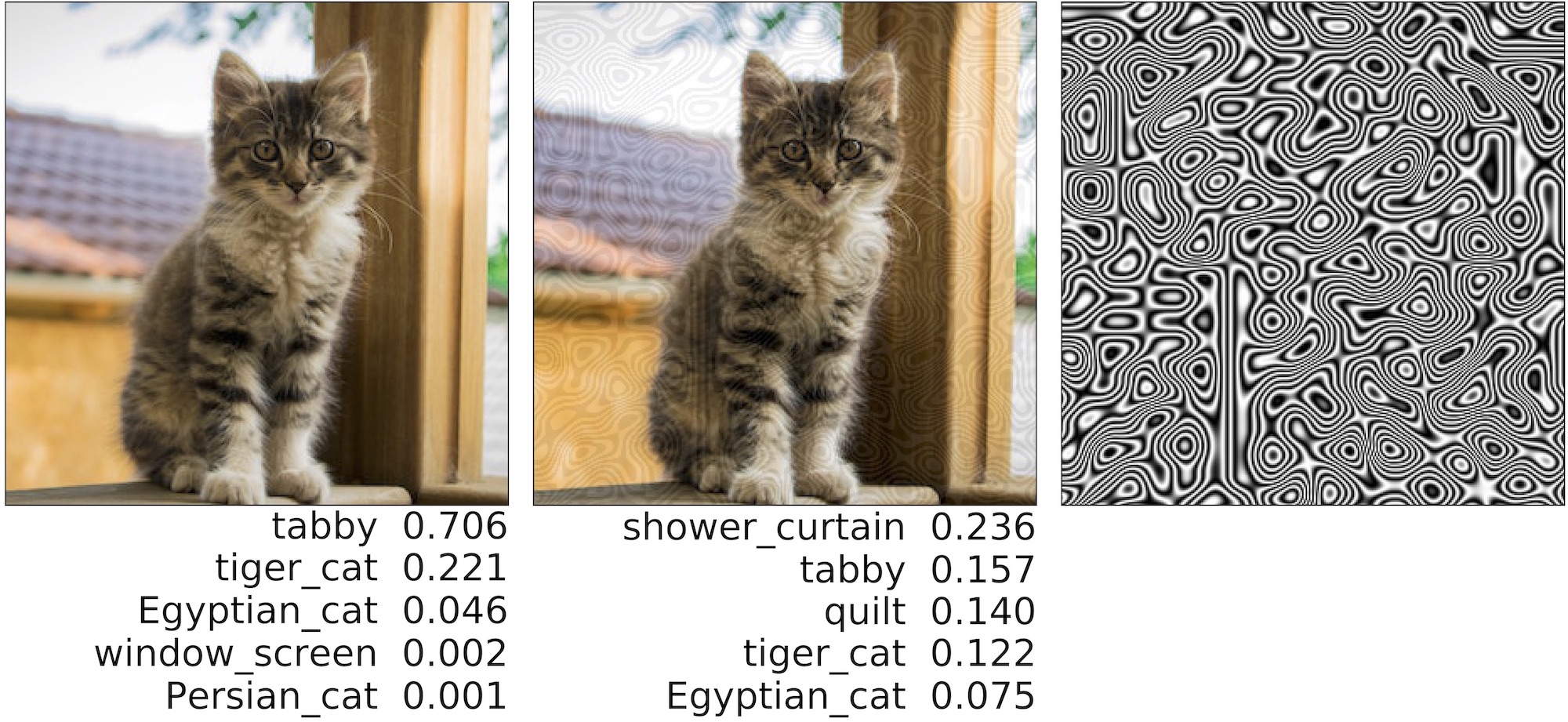}
	\caption{Adversarial example generated with a procedural noise function. From left to right: original image, adversarial example, and procedural noise (magnified for visibility). Below are the classifier's top 5 output probabilities.}
	\label{fig:1_adv_example}
\end{figure}

It has been shown that machine learning systems are vulnerable to attacks performed at test time \cite{barreno2006, huang, mcdaniel, papernot2018sok, luisSecret}. In particular, DCNs have been shown to be susceptible to \textit{adversarial examples}: inputs indistinguishable from genuine data points but designed to be misclassified by the learning algorithm \cite{szegedy2013intriguing}. As the perturbation required to fool the learning algorithm is usually small, detecting adversarial examples is a challenging task. Fig.~\ref{fig:1_adv_example} shows an adversarial example generated with the attack strategy we propose in this paper; the perturbed image of a tabby cat is misclassified as a shower curtain. Although we focus on computer vision, this phenomenon has been shown in other application domains such as speech processing \cite{carlini2018audio, carlini2016hidden}, malware classification \cite{grosse2016adversarial}, and reinforcement learning \cite{huang2017adversarial, lin2017tactics} among others.





In this paper, we propose a novel approach for generating adversarial examples based on the use of procedural noise functions. Such functions are commonly used in computer graphics and designed to be parametrizable, fast, and lightweight \cite{lagae2010survey}. Their primary purpose is to algorithmically generate textures and patterns on the fly. Procedurally generated noise patterns have interesting  structures that are visually similar to those in existing universal adversarial perturbations \cite{moosavi2017universal, khrulkov2018art}.

We empirically demonstrate that DCNs are fragile to procedural noise and these act as \emph{Universal Adversarial Perturbations (UAPs)}, i.e. input-agnostic adversarial perturbations. Our experimental results on the large-scale \emph{ImageNet} classifiers show that our proposed black-box attacks can fool classifiers on up to 98.3\% of input examples. The attack also transfers to the object detection task, showing that it has an obfuscating effect on objects against the YOLO v3 object detector \cite{redmon2018yolov3}. These results suggest that large-scale indiscriminate black-box attacks against DCN-based machine learning services are not only possible but can be realized at low computational costs. Our contributions are as follows:
\begin{itemize}
\item We show a novel and intuitive vulnerability of DCNs in computer vision tasks to procedural noise perturbations. These functions characterize a distribution of noise patterns with high universal evasion, and universal perturbations optimized on small datasets generalize to datasets that are 10 to 100 times larger. To our knowledge, this is the first model-agnostic black-box generation of universal adversarial perturbations.
\item We propose \emph{Bayesian optimization} \cite{mockus1978, snoek2012practical} as an effective tool to augment black-box attacks. In particular, we show that it can use our procedural noise to craft inexpensive universal and input-specific black-box attacks. It improves on the query efficiency of random parameter selection by 5-fold and consistently outperforms the popular L-BFGS optimization algorithm. Against existing query-efficient black-box attacks, we achieve a 100 times improvement on the query efficiency while maintaining a competitive success rate.
\item We show evidence that our procedural noise UAPs appear to exploit low-level features in DCNs, and that this vulnerability may be exploited to create universal adversarial perturbations across applications. We also highlight the shortcomings of adversarial training and suggest using more input-agnostic defences to reduce model sensitivity to adversarial perturbations.
\end{itemize}

The rest of the paper is structured as follows. In Sect.~2, we define a taxonomy to evaluate evasion attacks. In Sect.~3, we describe and motivate the use of procedural noise functions. In Sect.~4, we demonstrate how different DCN architectures used in image classification have vulnerabilities to procedural noise. In Sect.~5, we show how to leverage this vulnerability to create efficient black-box attacks. In Sect.~6, we analyze how the attack transfers to the object detection task and discuss how it can generalize to other application domains. In Sect.~7, we explore denoising as a preliminary countermeasure. Finally, in Sect.~8, we summarize our findings and suggest future research directions.

\section{Attack Taxonomy}
Our study focuses on attacks at test time, also known as \textit{evasion attacks}. To determine the viability and impact of attacks in practical settings, we categorize them according to three factors: (a) the generalizability of their perturbations, (b) the access and knowledge the adversary requires, and (c) the desired output. These factors also describe the threat model being considered.

\subsection{Generalizability}
The \textit{generalizability} of adversarial perturbations refers to their ability to apply across a dataset or to other models. Perturbations that generalize are more efficient because they do not need to be re-computed for new data points or models. Their generalizability can be described by their \textit{transferability} and \textit{universality}.

\textit{Input-specific} adversarial perturbations are designed for a specific input against a given model, these are neither transferable or universal. \textit{Transferable} adversarial perturbations can fool multiple models \cite{papernot2016transferability} when applied to the same input. This property enhances the strength of the attack, as the same adversarial input can degrade the performance of multiple models, and makes possible black-box attacks through surrogate models. Perturbations are \textit{universal} when the same adversarial perturbation can be applied successfully across a large portion of the input dataset to fool a classifier \cite{moosavi2017universal}. \emph{Cross-model universal} perturbations are both transferable and universal, i.e., they generalize across both a large portion of the inputs and across models. Generating adversarial perturbations that generalize is suitable and more efficient in attacks that target a large number of data points and models, i.e. for broad spectrum indiscriminate attacks. In contrast, input-specific attacks may be easier to craft when a few specific data points or models are targeted or for targeted attacks where the attacker aims to produce some specific types of errors.

\subsection{Degree of Knowledge}
For evasion attacks, we assume that the attacker has access to the test input and output. Beyond this, the adversary's knowledge and capabilities range from no access or knowledge of the targeted system to complete control of the target model. Accordingly, attacks can be broadly classified as: white-box, grey-box, or black-box \cite{papernot2018sok}.

In \textit{white-box} settings, the adversary has complete knowledge of the model architecture, parameters, and training data. This is the setting adopted by many existing studies including \cite{szegedy2013intriguing, carlini2017towards, goodfellow2014explaining, kurakin2016adversarial, madry2017towards, moosavi2016deepfool}. In \textit{grey-box} settings, the adversary can build a surrogate model of similar scale and has access to training data similar to that used to train the target system. This setting is adopted in transfer attacks where white-box adversarial examples are generated on a surrogate model to attack the targeted model \cite{papernot2017practical, kurakin2016scale}. This approach can also be adapted for a black-box setting. For example Papernot et al. \cite{papernot2017practical} apply a heuristic to generate synthetic data based on queries to the target classifier, thus removing the requirement for labelled training data. In a \textit{black-box} setting, the adversary has no knowledge of the target model and no access to surrogate datasets. The only interaction with the target model is by querying it, this is often referred to as an ``oracle''.


Given the adversary's lack of knowledge, black-box attacks rely heavily on making numerous queries to gain information. This increases the chances of the attack being detected. Thus, the most dangerous attacks are those that require the least queries and resources. With fewer queries, adversarial perturbations are generated sooner, costs (when using a paid service) are lower and the volume of suspicious queries is reduced. Existing black box attacks like \cite{bhagoji2018black, chen2017zoo} have shown some success with zeroth order optimization and gradient estimation. However they require tens to hundreds of thousands of queries on datasets with a large number of features, as in realistic natural-image dataset like ImageNet \cite{du2018towards, ilyas2018prior}. The most query-efficient method reported so far is a bandit optimization framework that achieves 92.9\%  success with an average of $1,000$ queries per image on ImageNet \cite{ilyas2018prior}.

\subsection{Desired Output}
The adversary's goal varies according to the application and the expected rewards gained from exploiting the system. Usually, attacks are considered as either \textit{targeted} or \textit{untargeted (indiscriminate)}.

In \textit{targeted} attacks the adversary aims for a specific subset of inputs to be misclassified as their chosen output. In \textit{untargeted} attacks, the attacker aims to cause classification errors on a subset of inputs. Both these attacks disrupt the machine learning system by forcing errors and undermining the model's reliability.


In multi-class classification, targeted attacks are more challenging due to their specificity, but successful ones allow a greater degree of manipulation for the attacker. On the other side, untargeted attacks are typically easier, as they just have to evade the ``correct'' classification, and this characteristic is more suited for broad indiscriminate attacks.

\section{Procedural Noise}
We introduce \textit{procedural noise functions} as an intuitive and computationally efficient approach to generate adversarial examples in black-box settings. Procedural noise functions are algorithmic techniques used for generating image patterns, typically used in the creation of natural details to enhance images in video and graphics production. They are designed to be fast to evaluate, scale to large dimensions, and have low memory footprint. These attributes make them desirable for generating computationally inexpensive perturbations. For a more comprehensive survey on procedural noise, we refer the reader to \citet{lagae2010survey}.

\subsection{Motivation}
Existing \textit{Universal Adversarial Perturbations (UAPs)} generated by white-box attacks exhibit interesting visual structures that are \cite{moosavi2017universal}, as of yet, not fully understood. UAPs are particularly interesting as their universality reveals more generic or class-agnostic features that machine learning algorithms appear to be sensitive to. In contrast, input-specific adversarial perturbations, though less detectable in many cases, can ``overfit'' and apply only to the inputs they were designed for \cite{zhou2018transferable}.

Current approaches, like the following, craft UAPs using white-box knowledge of the model's learned parameters. \citet{moosavi2017universal} use the DeepFool algorithm \cite{moosavi2016deepfool} iteratively over a set of images. \citet{mopuri2018nag} use Generative Adversarial Nets (GANs) to compute UAPs, whilst \citet{khrulkov2018art} propose to use the singular vector method that maximizes the difference in activations at a targeted hidden layer between the original and the adversarial examples.


We hypothesize that procedural noise, which exhibits patterns visually similar to those of UAPs (see Appendix~\ref{appendix:ex_perturbs}), can also act as a UAP. Procedural noise is simple to implement, fast to compute, and does not require the additional overhead of building, training, or accessing a DCN to generate adversarial perturbations. The parametrization of procedural noise is simpler and this results in a reduced search space, which can enable query-efficient black-box attacks. This is particularly useful in large-scale applications like natural-image classification where existing attacks explore the entire input space, which has very high dimensionality ($\ge$ 100,000). 

Procedural noise functions can be classified into three categories: lattice gradient noise, sparse convolution noise, and explicit noise \cite{lagae2010survey}. Lattice gradient noise is generated by interpolating random values or gradients at the points of an integer lattice, with \textbf{Perlin noise} as a representative example. Sparse convolution noise is a sum of randomly positioned and weighted kernels,\footnote{A kernel in image processing refers to a matrix used for image convolution.} with \textbf{Gabor noise} as a representative example. Explicit noise differs from the others in that the images are generated in advance and stored later for retrieval. This induces large memory costs and limits its applicability as an inexpensive attack. We therefore do not use it here and leave its investigation for future work.

\subsection{Perlin Noise}
We chose to use Perlin noise as a representative example for lattice gradient noise, because of its ease of use, popularity, and simplicity. Perlin noise was developed as a technique to produce natural-looking textures for computer graphics, with its initial application in motion pictures where it has remained a staple of the industry \cite{lagae2010survey}. Although Perlin noise may not be the most expressive noise function, it has a simple implementation which makes it suitable for inexpensive black-box attacks as it is controlled by only a few parameters.

We summarize the formal construction of two-dimensional Perlin nose as described by Perlin \cite{perlin2002improvingcode, perlin2002improving, perlin1985image}. The value at a point $(x, y)$ is derived as follows: let $(i, j)$ define the four lattice points of the lattice square where $i = \{ |x|, |x| + 1 \}$ and $j = \{ |y|, |y| + 1 \}$. The four gradients are given by $q_{ij} = \mathbf{V}[\mathbf{Q}[\mathbf{Q}[i] + j]]$ where precomputed arrays $\mathbf{Q}$ and $\mathbf{V}$ contain a pseudo-random permutation and pseudo-random unit gradient vectors respectively. The four linear functions $q_{ij} (x - i, y - j)$ are then bilinearly interpolated by $s(x - |x|)$ and $s(y - |y|)$, where $s(t) = 6 t^5 - 15 t^4 + 10 t^3$. The result is the Perlin noise value $p(x, y)$ for coordinates $(x, y)$.

The Perlin noise function has several parameters that determine the visual appearance of the noise. In our implementation, the wavelengths $\lambda_x, \lambda_y$ and number of octaves $\Omega$ contribute the most to the visual change. The noise value at point $(x, y)$ with parameters $\delta_{\text{per}} = \{\lambda_x, \lambda_y, \Omega\}$ becomes
$$ S_{\text{per}} (x, y) = \sum_{n = 1}^\Omega p(x \cdot \tfrac{2^{n - 1}}{\lambda_x}, y \cdot \tfrac{2^{n - 1}}{\lambda_y}) $$

To achieve more distinct visual patterns, we use a sine colour map with an additional frequency parameter $\phi_{\text{sine}}$. This colour map for the noise value $p$ is defined by $C(p) = \sin(p \cdot 2 \pi \phi_{\text{sine}})$. The periodicity of the sine function creates distinct bands in the image to achieve a high frequency of edges. The resulting noise generating function $G_{\text{per}}$ at point $(x, y)$ is defined as the composition of our Perlin noise function and the sine colour map,
$$ G_{\text{per}} (x, y) = C(S_{\text{per}} (x, y)) = \sin((S_{\text{per}} (x, y) \cdot 2 \pi \phi_{\text{sine}}) $$
with combined parameters $\delta_{\text{per}} = \{\lambda_x, \lambda_y, \phi_{\text{sine}}, \Omega\}$.

\subsection{Gabor Noise}
We use Gabor noise as a representative example for sparse convolution noise. It has more accurate spectral control than other procedural noise functions \cite{lagae2009procedural}, where spectral control refers to the ability to control the appearance of noise as measured by its energy along each frequency band. Gabor noise can be quickly evaluated at any point in space and is characterized by only a few parameters such as orientation, frequency, and bandwidth \cite{lagae2009procedural}. In essence, Gabor noise is a convolution between sparse white noise and a Gabor kernel $g$. The Gabor kernel is the product of a circular Gaussian and a Harmonic function:
$$ g(x, y) = e^{-\pi \sigma^2 (x^2 + y^2)} \cos{[\tfrac{2 \pi}{\lambda} (x \cos \omega + y \sin \omega)]}, $$
where $\sigma$ is the width of the Gaussian, $\lambda$ and $\omega$ are the period and orientation of the Harmonic function \cite{lagae2010survey}. Our implementation of the Gabor kernel uses the OpenCV library \cite{opencv_library}. The value $S_{\text{gab}}(x, y)$ at point $(x, y)$ is the sparse convolution with a Gabor kernel where $\{(x_i, y_i)\}$ are the random points \cite{lagae2010survey}. 
Gabor noise is an expressive noise function and will have a large number of dimensions if each random point is assigned different kernel parameters. To simplify the implementation, we use the same parameters and weights for each random point $(x_i, y_i)$. This results in noise patterns that have a uniform texture. 

We add an additional discrete parameter $\xi$ to control the isotropy of the Gabor noise. Having $\xi = 1$ results in anisotropic noise, which is oriented in one direction. Progressively larger $\xi$ makes it more uniform in all directions. Implementing these changes gives an updated formulation
$$ S_{\text{gab}} (x, y) = \tfrac{1}{\xi} \sum_{n = 1}^\xi \sum_i g(x - x_i, y - y_i; \sigma, \lambda, \omega + \tfrac{n \pi}{\xi}). $$

To achieve high-frequency patterns and remove unwanted low-contrast oscillations, we normalize the variance spectrum of the Gabor noise using the algorithm described by \citet{neyret2016understanding}. This results in min-max oscillations. 
Note that we refer specifically to the \textit{frequency of edges}, i.e. how often edges appear one after the other on an image. This is different from the classical frequency used in spectral analysis. For simplicity we use ``low-frequency'' or ``high-frequency'' when referring to patterns with low or high edge frequency in the image.

We have tried applying the sine colour map to Gabor noise, but it obscures the Gabor kernel structures. Similarly, we have tried normalizing the variance spectrum of Perlin noise, but this resulted in flat images with few edges and no distinct patterns. The final noise generating function $G_{\text{per}}$ with a normalized variance spectrum has parameters $\delta_{\text{gab}} = \{\sigma, \lambda, \omega, \xi\}$.

\section{Vulnerability to Procedural Noise}
In this section, we empirically demonstrate the vulnerability of five different DCN architectures to procedural noise for the ImageNet classification task. We show that randomly chosen procedural noise perturbations act as effective UAPs against these classifiers. The procedural noise UAPs greatly exceed the baseline uniform random noise, is on average universal on more than half the dataset across different models, and fools each model on more than 72\% of the dataset when considering input-specific evasion.

\subsection{Experiment Setup}
The data used are 5,000 random images from the validation set of the ILSVRC2012 ImageNet classification task \cite{russakovsky2015imagenet}. It is a widely used object recognition benchmark with images taken from various search engines and manually labelled to 1,000 distinct object categories where each image is assigned one ground truth label.

\textbf{Models.} We use four distinct DCN architectures pre-trained on ImageNet: VGG-19 \cite{simonyan2014very}, ResNet-50 \cite{he2016deep}, Inception v3 \cite{szegedy2016rethinking}, and Inception ResNet v2 \cite{szegedy2017inception}. We abbreviate Inception ResNet v2 to \textbf{IRv2}. Inception v3 and Inception ResNet v2 take input images with dimensions $299 \times 299 \times 3$ while the remaining two networks take images with dimensions $224 \times 224 \times 3$.

We also take an ensemble adversarially trained version of the Inception ResNet v2 architecture: Tramer et al. \cite{tramer2017ensemble} fine-tuned the IRv2 network by applying ensemble adversarial training, making it more robust to gradient-based attacks. We refer to this new model as \textbf{IRv2$_{\text{ens}}$}. It has the same architecture, but different weights when compared to the first IRv2. For complete details of the ensemble adversarial training process, we refer the reader to \cite{tramer2017ensemble}.

\textbf{Perturbations.} The Perlin noise parameters $\lambda_x, \lambda_y, \phi_{\text{sine}}$ and Gabor noise parameters $\sigma, \lambda$ are positive and bounded above by the image's side length $d$. Increasing the parameters beyond this will have no impact on the resulting image as it gets clipped by the image dimension. The isotropy $\xi \in [1, 12]$ and number of octaves $\Omega \in [1, 4]$ are discrete, and evaluations show negligible change in the image for $\xi > 12$ and $\Omega > 4$. The range on the angle $\omega \in [0, 2\pi]$ covers all directions.

We fix the kernel size and number of random points for Gabor noise so that the Gabor kernels always populate the entire image. The parameters $\delta_{\text{gab}}$ of the Gabor noise have the greater influence on the resulting visual appearance of the noise pattern.

To provide a baseline, we also test the models against uniform random noise perturbations: $\text{sgn}(r) \cdot \varepsilon$ where $r \in \mathcal{U}(-1, 1)^{d \times d \times 3}$, $d$ is the image's side length, and $\varepsilon$ is the $\ell_{\infty}$-norm constraint on the perturbation. This is an $\ell_{\infty}$-optimized uniform random noise, and it is reasonable to say that attacks that significantly outperform this baseline are non-trivial.

\textbf{Metrics.} To measure the universality of a perturbation, we define the \emph{universal evasion rate} of a perturbation over the dataset. Given model output $f$, input $x \in X$, perturbation $s$, and small $\varepsilon > 0$, the universal evasion of $s$ over $X$ is
$$ \frac{\vert \{x \in X : \argmax f(x + s) \neq \tau(x) \} \vert}{\vert X \vert}, \quad \Vert s \Vert_{\infty} \leq \varepsilon, $$
where $\tau(x)$ is the true class label of $x$. An $\ell_{\infty}$-norm constraint on $s$ ensures that the perturbation is small and does not drastically alter the visual appearance of the resulting image; this is a \emph{proxy} for the constraint $\tau(x) = \tau(x + s)$. We choose the $\ell_{\infty}$-norm as it is straightforward to impose for procedural noise perturbations and is often used in the adversarial machine learning literature. In this case, $f$ is the probability output vector, but note that the attacker only needs to know the output class label $\argmax f(x + s)$ to determine if their attacks succeed. To measure the model's sensitivity against the perturbations for each input, we define the model's \emph{average sensitivity} on an input $x$ over perturbations $s \in S$ as
$$ \frac{\vert \{s \in S : \argmax f(x + s) \neq \tau(x) \} \vert}{\vert S \vert}, \quad \Vert s \Vert_{\infty} \leq \varepsilon. $$
This will help determine the portion of the dataset on which the model is more vulnerable. Finally, the \emph{input-specific evasion} rate is
$$ \frac{\vert \{x \in X : \exists s \in S \text{ such that } \argmax f(x + s) \neq \tau(x) \} \vert}{\vert X \vert}, \quad \Vert s \Vert_{\infty} \leq \varepsilon. $$
This measures how many inputs in the dataset can be evaded with perturbations from $S$.

\textbf{Experiment.} We evaluate our procedural noise perturbations on 5,000 random images from the validation set. The perturbations generated are from 1,000 Gabor noise, 1,000 Perlin noise, and 10,000 uniform random perturbations. 1,000 queries for the procedural noise functions was sufficient for our results and its search space only has four bounded parameters. We use an $\ell_{\infty}$-norm constraint of $\varepsilon = 16$. The pixels of the perturbations are first clipped to $[-\varepsilon, \varepsilon]$ and the resulting adversarial example's pixels are clipped to the image space $[0, 255]$.

Thus, this is an untargeted black-box attack with exactly 1,000 queries per image for procedural noise. Note that the adversary has no knowledge of the target model and only requires the top label from the model's output.

\begin{figure*}[t]
    \centering
    \begin{subfigure}[t]{\textwidth}
        \centering
        \includegraphics[trim = {0 0.2in 0 0}, width = \textwidth]{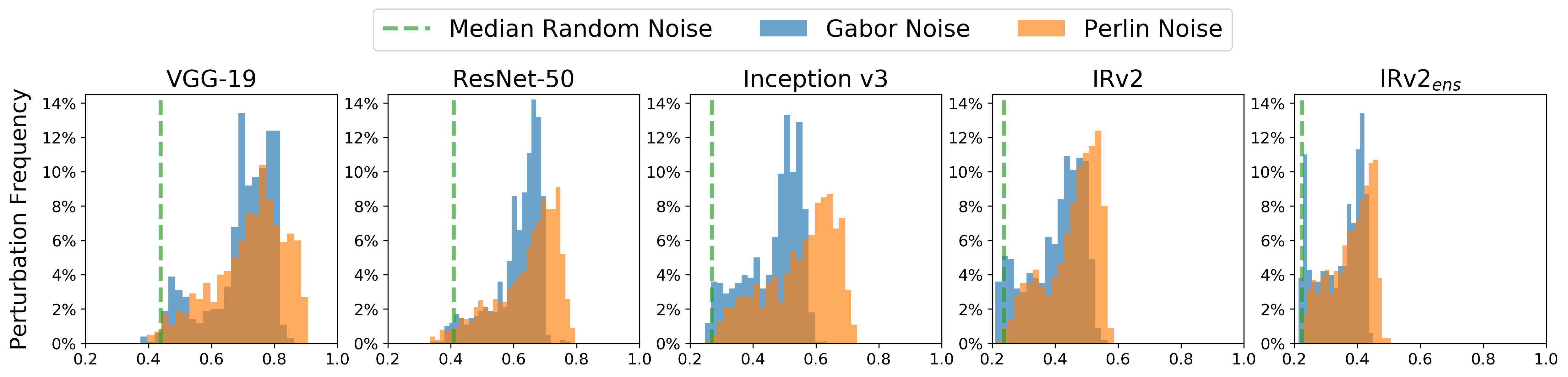}
        \caption{Universal Evasion Rate}
        \label{fig:4_histogram_univ}
    \end{subfigure}%
    \\
    \begin{subfigure}[t]{\textwidth}
        \centering
        \includegraphics[trim = {0 0.2in 0 -0.2in}, width = \textwidth]{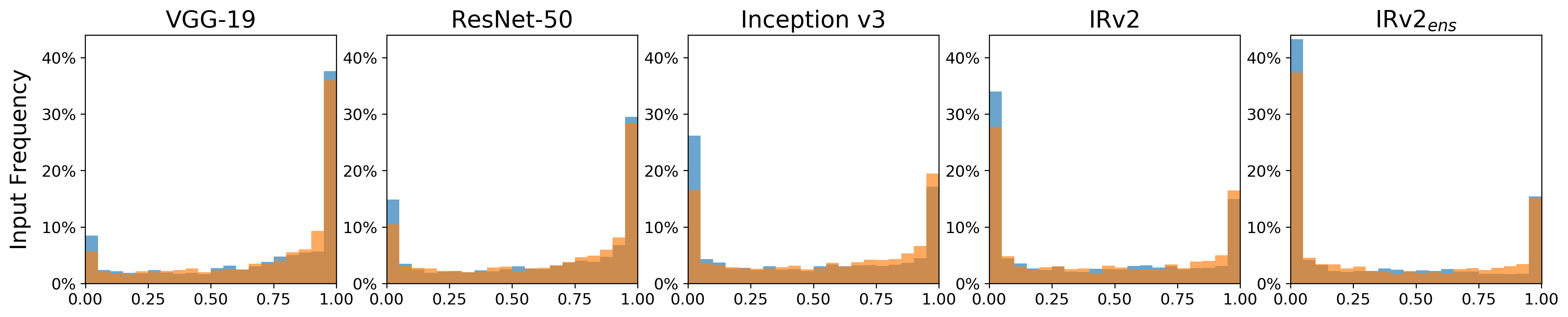}
        \caption{Average Sensitivity}
        \label{fig:4_histogram_sens}
    \end{subfigure}
    \caption{Histogram of (a) universal evasion rate over all perturbations and (b) average sensitivity over all inputs across all models. For example, VGG-19 in (a) shows that about 12\% of the Gabor noise perturbations have a universal evasion rate of approximately 0.8.}
    \label{fig:4_histogram}
\end{figure*}

\subsection{Universality of Perturbations}
The procedural noise perturbations appear to be both universal and transferable across models. The results in Fig.~\ref{fig:4_histogram} show that the models in order from least to most robust against the perturbations are: VGG-19, ResNet-50, Inception v3, IRv2, and then IRv2$_{\text{ens}}$. This is not surprising, as the generalization error for these models appear in the same order, with larger generalization error indicating a less robust model. The ensemble adversarially trained model has also been hardened against adversarial examples, so it was expected to be less affected. However, the ensemble adversarial training did not fully mitigate the impact of the procedural noise.

Both Gabor and Perlin noise have significantly higher universal evasion rate than random noise. In Fig.~\ref{fig:4_histogram_univ}, we represent random noise using the median (which, in this case, is very close to the mean) of its universal evasion rate over all 10,000 perturbations, as the variance is very small (less than $10^{-5}$) for each model.

Procedural noise functions create a distribution whose modes have high universal evasion rates. Table~\ref{table:4_quartiles_in3} shows an example of this on Inception v3, where more than half the Perlin noise perturbations achieve evasion on more than 57\% of the dataset.

Between the two procedural noises, Perlin noise is a stronger UAP, though not by a large margin. In Fig.~\ref{fig:4_histogram_univ}, Gabor noise appears bimodal, especially with the IRv2 classifiers. This bimodal aspect shows that there are two large disjoint subsets of Gabor noise: one with high universal evasion and another with low universal evasion. This was not as prominent for Perlin noise.

\begin{table}[htp]
	\caption{Universal evasion (\%) by percentile for random and procedural noise perturbations on Inception v3.}
	\label{table:4_quartiles_in3}
	\small
\begin{tabular*}{2.3in}{@{\extracolsep{\fill}} l ccc @{}}
	Percentile & Random & Gabor & Perlin\\
	\midrule
	min	& 26.0 & 24.7 & \textbf{26.4}\\
	25th	& 26.8 & 39.9 & \textbf{47.2}\\
	50th	& 27.0 & 49.1 & \textbf{57.7}\\
	75th	& 27.1 & 53.3 & \textbf{63.7}\\
	max	& 27.9 & 61.8 & \textbf{73.2}\\
	\bottomrule
\end{tabular*}
\end{table}

\textbf{Best Parameters.} Given the distribution of universal evasion rates in Fig.~\ref{fig:4_histogram_univ}, it is interesting to observe which parameters of the noise functions contribute most to the evasion rate. Because there are only four parameters for each procedural noise function, the linear correlations between these parameters and the universal evasion rates give a general overview of their influence. Though this analysis may not fully capture multi-variable relationships.

For Gabor noise, the parameters $\sigma$, $\omega$, and $\xi$ have low correlation ($\le 0.16$) with universal evasion. The remaining parameter $\lambda$ however has a high correlation ($\ge 0.52$) with the universal evasion rate across all classifiers; this peaks at 0.90 for IRv2$_{\text{ens}}$. $\lambda$ corresponds to the wavelength of the harmonic kernel. Visually, lower values of $\lambda$ indicate thinner and more frequent bands, which is why $\lambda$ can also be thought of as an inverse frequency. The correlations suggest that low-frequency Gabor noise patterns correlate with high universal evasion, and this appears to be more pronounced as we move to the more robust models. 

For Perlin noise, the parameters $\lambda_x$ and $\lambda_y$ have low correlation ($\le 0.25$) with universal evasion. The universal evasion rates have a moderate negative correlation with the number of octaves $\Omega$ and have a moderately-high positive correlation with $\phi_{\text{sine}}$ for the non-adversarially trained models. This suggests that high-frequency Perlin noise patterns correlate with high universal evasion. The number of octaves indicate the amount of detail or the number of curvatures within the image, so a slight negative correlation indicates that some patterns that achieve high evasion have fewer curvatures and details. Visually, the frequency of the sine function $\phi_{\text{sine}}$ can be thought of as the thinness of the bands in the image. This is the opposite of the $\lambda$ in Gabor noise, so we have a similar result where IRv2$_{\text{ens}}$ is more susceptible to low-frequency patterns.

Low-frequency patterns are also more effective on IRv2$_{\text{ens}}$ because this model was adversarially trained on gradient-based attacks, which we hypothesize generated mostly \emph{high-frequency} perturbations. The frequency appears to be an important factor in determining the strength of the UAP even when Gabor and Perlin noise have opposite correlations between their frequency and their universal evasion. This difference shows that these noise patterns have notably different characteristics. Complete correlation matrices are available in Appendix~\ref{appendix:corr_mat}.


\textbf{Cross-model Universality.} The UAPs with high universal evasion on one model often have high universal evasion on the other models. For example, a Perlin noise perturbation has 86.7\%, 77.4\%, 73.2\%, 58.2\%, and 45.9\% universal evasion on VGG-19, ResNet-50, Inception v3, IRv2, and IRv2$_{\text{ens}}$ respectively.

The correlation of the universal evasion rates across models is high ($\ge 0.80$), which shows the transferability of our UAPs across models. For Gabor noise, there is a high correlation across all models. For Perlin noise, there appears to be a noticeable difference between IRv2$_{\text{ens}}$ and the remaining four models. The strongest Perlin noise on the undefended models have high-frequency patterns that the adversarial training partially mitigated.

Overall, the results show that procedural noise perturbations act as UAPs across all models. This is the first black-box generation of cross-model UAPs, and the procedural noise allows us to draw perturbations from a distribution with high average universal evasion. We now look at the results from an input-specific perspective.

\subsection{Model Sensitivity on Inputs}
The model's sensitivity could vary across the input dataset, i.e. the model's predictions are stable for some inputs while others are more susceptible to small perturbations. We measure this with the average sensitivity of single inputs over all perturbations. Results in Fig.~\ref{fig:4_histogram_sens} show that the average sensitivity of the dataset is bimodal for all models and both procedural noise functions. There are two distinct subsets of the data: one on the right that is very sensitive and the other on the left that is very insensitive to the perturbations. The remaining data points are somewhat uniformly spread in the middle. The number of points on the left peak of the histogram in Fig.~\ref{fig:4_histogram_sens} is larger for the most robust models. Similarly to Fig.~\ref{fig:4_histogram_univ}, this progression indicates that the order of most to least sensitive models align with the most to least robust models. We omit results for random noise since more than 60\% of the input dataset are barely affected by it across all models. This manifests as a tall peak on the left and a short peak on the right.

When comparing the average sensitivity of inputs between the two procedural noise functions on the same models, the correlations range between 0.89-0.92, which shows that both procedural noise perturbations affect very similar groups of inputs for each model. The correlation between the average sensitivities for each input across the Inception models is at least 0.79, which suggests that these models are sensitive to procedural noise on similar inputs. This is less so between ResNet-50 and VGG-19 whose correlations with the other models range from 0.56-0.81.

\textbf{Input-specific Evasion.} We consider the case when our untargeted black-box attack is used as an input-specific attack, i.e. the adversary only needs to find at least one adversarial perturbation that evades each input. Thus, Evasion on input $x$ is achieved when $\exists s \in S$ such that $\argmax f(x + s) \neq \tau(x)$. This is in contrast to the universal attack where the adversary crafts a single perturbation to fool the model on as many inputs as possible.

Note that the random noise is optimized for the $\ell_{\infty}$-norm. We draw the random noise perturbation from $\{-\varepsilon, \varepsilon\}^{d \times d \times 3}$. Thus, for each pixel the added noise is either $-\varepsilon$ or $\varepsilon$, rather than drawing from the continuous domain $(-\varepsilon, \varepsilon)$. It is reasonable to think that a larger perturbation is more likely to cause evasion.

Table~\ref{table:4_input-spec} shows that VGG-19 and ResNet-50 are particularly fragile as even random noise greatly degrades their performance. Although the Inception models are more robust, both procedural noise perturbations still achieve more than 72\% evasion on all of them. Although ensemble adversarial training improved the robustness of IRv2, it still does not mitigate the impact of the procedural noise attack. The idea of ensemble adversarial training was to decouple the generation of the adversarial training set from the original model, but this was limited since it was only done for gradient-based attacks. We argue that defences should be more input-agnostic to avoid having to train against all types of attacks.

\begin{table}[htb]
	\caption{Input-specific evasion rate (in \%) for random and procedural noise perturbations. Original refers to the top 1 error on the unaltered original images. Strongest attack on each classifier is highlighted.}
	\label{table:4_input-spec}
	\small
	\begin{tabular*}{3in}{@{\extracolsep{\fill}} l cccc @{}}
		Classifier & Original & Random & Gabor & Perlin\\
		\midrule
		VGG-19			& 29.4 & 57.1 &	 97.7 & \textbf{98.3}\\
		ResNet-50		& 25.9 & 55.7 & 96.2 & \textbf{96.3}\\
		Inception v3		& 22.3 & 46.8 &	 89.2 & \textbf{93.6}\\
		IRv2				& 20.1 & 38.7 & 81.1 & \textbf{87.0}\\
		IRv2$_{\text{ens}}$	& 20.1 & 37.5 & 72.7 & \textbf{79.4}\\
		\bottomrule
	\end{tabular*}
\end{table}

\textbf{Label Analysis.} The procedural noise perturbations were not designed to be a targeted attack, but intuitively, the same universal perturbation would cause misclassification towards class labels that have visually similar textures to the procedural noise pattern. We find that this holds true for only a few of the procedural noise UAPs.

For a given procedural noise perturbation, we define its \emph{top label} to be the class label it causes the most misclassification towards. When looking at procedural noise UAPs across all models, about 90\% of Gabor noise perturbations have their top label apply to at most 9\% of the inputs. For Perlin noise, about 80\% of its perturbations have their top label on at most 10\% of the input. There are however a few outliers that have their top label appear above 10\% of the inputs. For example, on Inception v3, the top Gabor noise with 61.8\% universal evasion has ``window screen'' as its top label and it applies for 37.8\% of the evaded inputs. In contrast, another Gabor noise perturbation with 58.9\% universal evasion has ``quilt'' as its top label, but it only applies for 6.0\% of the evaded inputs. As a consequence, it is still possible to use procedural noise to create universal targeted UAPs aimed at specific class labels like ``window screen'' or ``brain coral''. However the class labels we can target is dependent on the procedural noise and the overall success of universal targeted attacks may be limited, as it is more difficult to make a perturbation both universal and targeted. 

Perlin noise has a relatively large amount of ``brain coral'' classifications, with other labels such as ``maze'' and ``shower curtain'' also appearing frequently in the top five most classified labels per classifier. For Gabor noise, there was no label that consistently appeared at the top across all models. These results indicate that Perlin noise has a larger bias towards certain classes, while Gabor noise is more indiscriminate.

\subsection{Discussion}
Adversarial examples exploit the model's sensitivity, causing large changes in the model's output by applying specific small changes to the input. More specifically, recent work has shown adversarial examples exploit "non-robust" features that the model learns but that are incomprehensible to humans \cite{ilyas2019adversarial}. It is likely that UAPs may be exploiting "universal" non-robust features that the DCN has learned.

\textbf{Textures.} Previous results have shown ImageNet-trained DCNs to be more reliant on textures rather than shapes \cite{geirhos2018imagenet}. Though not explicitly shown in the later Inception architectures (Inception v3, IRv2), these are still likely to have a strong texture-bias due to similarities in the training. The texture bias however does not fully explain why small, and sometimes imperceptible, changes to the texture can drastically alter the classification output. We attribute adversarial perturbations more to the sensitivity of the model. Moreover, we test our attack against an object detection model in Sect.~6, and show that it also degrades the performance of models that have a spatial component in their learning task.

\textbf{Generalizability of Procedural Noise.} From the label analysis of procedural noise UAPs, we observe that most perturbations with high universal evasion do not have a strong bias towards any particular class label---no particular class is targeted more than 10\% for over 80\% of the procedural noise UAPs. This suggest that UAPs leverage more generic low-level features that the model learns, which would explain their \emph{universality} and indiscriminate behaviour in causing misclassification.

Amongst white-box UAP attacks, our procedural noise has the closest visual appearance to perturbations from the Singular Vector Attack (SVA) by \citet{khrulkov2018art}. They generate UAPs targeting specific layers of DCNs, and found that targeting earlier layers of the network generated more successful UAPs. The patterns obtained for these earlier layers share a visual appearance with procedural noise. These layers also correspond to the low-level features learned by the network. Other evidence also suggests that procedural noise exploits low-level features, as feature visualization of earlier layers in neural networks share the same visual appearance with some procedural noise patterns. Convolutional layers induce a prior on DCNs to learn local spatial information \cite{goodfellow2016book}, and DCNs trained on natural-image datasets learn convolution filters that are similar in appearance to Gabor kernels and colour blobs \cite{yosinski2014transferable, olah2017feature}. Gabor noise appears to be a simple collection of low-level features whereas Perlin noise seems to be a more complex mixture of low-level features. This difference in the complexity of their visual appearance may explain why Perlin noise is a stronger attack than Gabor noise.

Procedural noise attacks transfer with high correlation across the models most likely because they share the same training set (ImageNet), learning algorithms (e.g. backpropagation), and have similar components in their architectures (e.g. convolutional layers). This increases the likelihood that they share input-agnostic vulnerabilities. In this way, our results appear to support the idea that DCNs are \emph{sensitive} to aggregations of low-level features.

\textbf{Security Implications.} In transfer learning, a model trained on one task is re-purposed as an initialization or fixed feature extractor for another similar or related task. When used as a feature extractor, the initial layers are often frozen to preserve the low-level features learned. The subsequent layers, closer to the output, are then re-trained for the new task \cite{yosinski2014transferable}. Hidden layers from models pre-trained on the ImageNet dataset are often re-used for other natural-image classification tasks \cite{oquab2014learning}. This makes it a notable target as vulnerabilities that exploit low-level features encoded in the earlier layers carry over to the new models.

Transfer learning has many benefits as training entire models from scratch for large-scale tasks like natural-image classification can be costly both computationally in training and in terms of gathering the required data. However, this creates a systemic threat, as subsequent models will also be vulnerable to attacks on low-level features like procedural noise. Precisely characterizing the extent of the vulnerability of re-purposed models to the same attack is an interesting direction for future work.

Procedural noise is an accessible and inexpensive way for generating UAPs against existing image classifiers. In our experiments the Gabor and Perlin noise functions modified only four parameters, and each parameter was bounded in a closed interval. Drawing from this space of perturbations has generated UAPs with high universal evasion rates. Attackers can take advantage of the small search space by using procedural noise to craft query-efficient untargeted black-box attacks as we will show in Sect.~5.

\textbf{Strengths \& Limitations.} Procedural noise is one of the first black-box generation of UAPs. Other existing black-box attacks often optimize for input-specific evasion, and other existing UAP generation methods are white-box or grey-box. Whilst procedural noise is also a generative model, it differs from other generative models like Bayesian networks, Generative Adversarial Networks (GANs), or Variational Autoencoders (VAEs) as it does not require the additional overhead of building and training these generative models--which often requires more resources and stronger adversaries to execute successfully.

Comparing procedural noise attacks with existing black-box attacks is not straightforward, as procedural noise naturally has high universal evasion rates. This gives procedural noise an advantage in that, despite having no access to the target model, randomly drawn procedural noise patterns are likely to have high universal evasion. The search space for our procedural noise has only four dimensions, whereas most attacks are designed for the whole input space of hundreds of thousands of dimensions. However, this does come at a cost, as procedural noise functions do not capture adversarial perturbations outside their codomain. Other attacks that explore the whole input space are able to take full advantage of a stronger adversary (i.e. white-box or grey-box setting) or larger query limits. We explore this trade-off further with input-specific black-box attacks in Sect.~5.3.

Another limitation of this experiment was the use of an $\ell_{\infty}$-norm constraint. Although it is often used in the literature as a proxy for human perception, $\ell_{p}$-norms have limitations \cite{goodfellow2018defense} e.g., low-frequency patterns appear to be more visible than high-frequency patterns for the same $\ell_{p}$-norm. This frequency could be used as an additional constraint and developing a more reliable proxy for human perception remains an interesting avenue for future work.

\textbf{Summary.} Procedural noise attacks specialize as untargeted black-box perturbations with naturally high universal evasion. Offensively, this has applications as a large-scale indiscriminate attack and, defensively, as a standard test on machine learning services. We expand on this perspective in the following sections. In Sect.~5, we develop query-efficient black-box attacks using procedural noise and Bayesian optimization. In Sect.~6, we apply our procedural noise attack against a DCN designed for object detection, showing that the attack can generalize to other tasks. In Sect.~7, we test an input-agnostic defence in median filter denoising.

\section{Efficient Black-box Attacks}
Whilst in previous sections we have shown that procedural noise functions are an efficient way to generate adversarial perturbations, another significant advantage they bring is their low-dimensional search space. This enables the use of query-efficient black-box optimization techniques that otherwise do not scale well to high-dimensional problems. In this section, we compare several black-box optimization techniques for both input-specific and universal attacks and show that Bayesian optimization is an efficient method for choosing parameters in such black-box attacks.

\subsection{Bayesian Optimization}
Bayesian optimization is a sequential optimization algorithm used to find optimal parameters for black-box objective functions \cite{mockus1978, snoek2012practical}. This technique is often effective in solving various problems with expensive cost functions such as hyperparameter tuning, reinforcement learning, and combinatorial optimization \cite{shahriari2016taking}. Bayesian optimization consists of a probabilistic surrogate model, usually a Gaussian Process (GP), and an acquisition function that guides its queries. GP regression is used to update the belief on the parameters with respect to the objective function after each query \cite{shahriari2016taking}.


\textbf{Gaussian Processes.} A GP is the generalization of Gaussian distributions to a distribution over functions and is typically used as the surrogate model for Bayesian optimization \cite{rasmussen2004gaussian}. We use GPs as they induce a posterior distribution over the objective function that is analytically tractable. This allows us to update our beliefs about the objective function after each iteration \cite{snoek2012practical}.

A Gaussian Process $\mathcal{GP}(m, k)$ is fully described by a prior mean function $m: \mathcal{X} \to \mathbb{R}$ and positive-definite kernel or covariance function $k: \mathcal{X} \times \mathcal{X} \to \mathbb{R}$. We describe GP regression to understand how our GP is updated when more observations are available. The following expressions give the GP prior and Gaussian likelihood respectively
\begin{align*}
p(f \mid \textbf{X}) &=\mathcal{N}(m(\textbf{X}), \textbf{K})\\
p(y \mid f, \textbf{X}) &= \mathcal{N}(f(\textbf{X}), \sigma^2 \textbf{I})
\end{align*}
where $\mathcal{N}$ denotes a normal distribution. Elements of the mean and covariance matrix are given by $m_i = m(\textbf{x}_i)$ and $K_{i, j} = k(\textbf{x}_i, \textbf{x}_j)$. Given observations $\{ \textbf{X}, \textbf{y} \}$ and an arbitrary point $\textbf{x}$, the updated posterior mean and covariance on the $n$-th query are given by
\begin{align*}
m_n(\textbf{x}) &=m(\textbf{x}) - (\textbf{K} + \sigma^2 \textbf{I})^{-1} (y - m(\textbf{X}))\\
k_n(\textbf{x}, \textbf{x}) &= k(\textbf{x}, \textbf{x}) - k(\textbf{x}, \textbf{X}) (\textbf{K} + \sigma^2 \textbf{I})^{-1} k(\textbf{X}, \textbf{x}).
\end{align*}

We take the mean function to be zero $m \equiv 0$ to simplify evaluation and since no prior knowledge can be incorporated into the mean function \cite{shahriari2016taking} as is the case for black-box settings. 
A GP's ability to model a rich distribution of functions rests on its covariance function which controls important properties of the function distribution such as differentiability, periodicity, and amplitude \cite{rasmussen2004gaussian, shahriari2016taking}. Any prior knowledge of the target function is encoded in the hyperparameters of the covariance function. In a black box setting, we adopt a more general covariance function in the Mat\'ern 5/2 kernel
$$k_{5/2}(\textbf{x}, \textbf{x}') = \left(1 + \frac{\sqrt{5}r}{l} + \frac{5r^2}{3l^2} \right) \exp{\left(- \frac{\sqrt{5}r}{l} \right)}$$
where $r = \textbf{x} - \textbf{x}'$ and $l$ is the length-scale parameter \cite{snoek2012practical}. This results in twice-differentiable functions, an assumption that corresponds to those made in popular black-box optimization algorithms like quasi-Newton methods \cite{snoek2012practical}.


\textbf{Acquisition Functions.} The second component in Bayesian optimization is an acquisition function that describes how optimal a query is. Intuitively, the acquisition function evaluates the utility of candidate points for the next evaluation \cite{brochu2010tutorial}. The two most popular choices are the \emph{Expected Improvement} (EI) and \emph{Upper Confidence Bound} (UCB) \cite{shahriari2016taking}. First we define $\mu(\textbf{x})$ and $\sigma^2(\textbf{x})$ as the predictive mean and variance of $g(\textbf{x})$ respectively. Let $\gamma(\textbf{x}) = \frac{g(\textbf{x}_\text{best}) - \mu(\textbf{x})}{\sigma(\textbf{x})}$. The acquisition functions are
\begin{align*}
\alpha_\text{EI}(\textbf{x}) &= \sigma(\textbf{x}) (\gamma(\textbf{x}) \Phi(\gamma(\textbf{x})) + \mathcal{N}(\gamma(\textbf{x}) \mid 0, 1))\\
\alpha_\text{UCB}(\textbf{x}) &= \mu(\textbf{x}) + \kappa \sigma(\textbf{x}), \quad \kappa > 0
\end{align*}
where $\Phi$ is the normal cumulative distribution function.
EI and UCB have both been shown to be effective and data-efficient in real black-box optimization problems \cite{snoek2012practical}. However, most studies have found that EI converges near-optimally and is better-behaved than UCB in the general case \cite{brochu2010tutorial, shahriari2016taking, snoek2012practical}. This makes EI the best candidate for our acquisition function.

\subsection{Universal Black-box Attack}
For a universal black-box attack using procedural noise, the goal is to find the optimal parameters $\delta^\ast$ for the procedural noise generating function $G$ so that the universal evasion rate of perturbation $G(\delta^\ast)$ generalizes to unknown inputs $X_\text{val}$. The attacker has a smaller set of inputs $X_\text{train}$ and optimizes their perturbation $\delta^\ast$ for this dataset. The performance of the attack is measured by its universal evasion rate over the validation set $X_\text{val}$. In a practical setting, this is where the attacker optimizes their procedural noise UAP over a small dataset, then injects that optimized perturbation to other inputs--with the goal of causing as many misclassifications as possible.

\textbf{Experiment.} We use the Inception v3 model, $\ell_{\infty}$-norm $\varepsilon = 16$, and the same 5,000 data points tested in Sect.~4 as $X_\text{val}$. $X_\text{train}$ are points from the ILSVRC2012 validation set not in $X_\text{val}$. We test for training set sizes of 50, 125, 250, and 500, which corresponds to 1\%, 2.5\%, 5\%, and 10\% of the validation set size.

We compare Bayesian optimization with Limited-memory BFGS (L-BFGS) \cite{liu1989limited}, a quasi-Newton optimization algorithm that is often used in black-box optimization and machine learning. As the procedural noise functions are non-differentiable, we estimate gradients using finite difference approximation. This gradient approximation is similar to what is used for other black-box attacks like zeroth-order optimization \cite{chen2017zoo}, but here it is applied to a significantly smaller search space. When L-BFGS converges, possibly to a local optima, we restart the optimization with a different random initial point, stopping when the query limit is reached and choosing the best optima value found.

We set a maximum query limit of 1,000 universal evasion evaluations on the training set $X_\text{train}$. In practice, this limit was not necessary as both algorithms converged faster to their optimal values: within the first 100 queries for Bayesian optimization and within the first 250 queries for L-BFGS. These are untargeted universal black-box attacks where the adversary has no knowledge of the target model and only requires the top label from the model's outputs.

\textbf{Results.} The best procedural noise UAP computed from the training sets generalized well to the much larger validation set, consistently reaching 70\% or more universal evasion on the validation set for Perlin noise. This is a surprising result as the training sets were 10-100 times smaller than the validation set. This may be due to the inherent universality of our procedural noise perturbations. We focus on Perlin noise as it outperforms Gabor noise, with the latter averaging 58\% universal evasion on the validation set.

Table~\ref{table:5_bb_uap} shows that Bayesian optimization (BayesOpt) reliably outperforms L-BFGS in terms of universal evasion rate on the training sets and the resulting universal evasion rates on the validation set. For comparison, random parameters for Perlin noise in Sect.~4 had a 98th percentile of 70.2\% and a maximum of 73.1\% universal evasion. Bayesian optimization consistently reached or passed this 98th percentile while L-BFGS did not. It is reasonable for these optimization algorithms not to beat the maximum since the training sets were significantly smaller. Similar trends appear between Bayesian optimization, random selection, and L-BFGS for Gabor noise perturbations. We include these results in the Appendix~\ref{appendix:bb}.

\begin{table}[htb]
	\caption{Comparison on Inception v3 between universal Perlin noise black-box attacks. Universal evasion rates (\%) of the optimized perturbations are shown for their respective training set and the validation set.}
\label{table:5_bb_uap}
	\small
\begin{tabular*}{2.6in}{@{\extracolsep{\fill}} l cccccccc @{}}
	\toprule
	& \multicolumn{2}{c}{BayesOpt$_{\text{per}}$} && \multicolumn{2}{c}{L-BFGS$_{\text{per}}$}\\
	\cmidrule{2-3} \cmidrule{5-6}
	Train Size & Train & Val. && Train & Val.\\
	\midrule
	50		& 78.0 & 71.4 && 74.0 & 69.9\\
	125		& 77.6 & 70.2 && 76.0 & 71.5\\
	250		& 71.6 & 71.2 && 71.2 & 69.7\\
	500		& 75.0 & \textbf{72.9} && 73.4 & 70.8\\
	\bottomrule
\end{tabular*}
\end{table}


\subsection{Input-specific Black-box Attack}
In this section, we use procedural noise for an input-specific black-box algorithm. The goal of the adversary is to evade as many inputs in the dataset, maximizing the input-specific evasion rate on inputs that are not misclassified by the model. An important metric here is the query-efficiency of these attacks, as requiring large volumes of queries \emph{per sample} becomes impractical in real-world scenarios.

\textbf{Metrics.} We define the \emph{success rate} of an attack to be its input-specific evasion excluding clean inputs that are already misclassified. The \emph{average queries} is measured over successful evasions.

\textbf{Experiment.} We use the Inception v3 model, $\ell_{\infty}$-norm $\varepsilon = 16$, and the same 5,000 data points from Sect.~4. For a given input $x$, the goal is to achieve evasion with $\argmax f(x + s) \neq \tau(x)$ by minimizing the probability of the true class label $\tau(x)$. In this case we allow the attacker to access the model's output probability vector, as the black-box optimization algorithms gain minimal usable information if the output is binary ($\tau(x)$ or $\neg \tau(x)$).

As in the previous section, we compare Bayesian optimization and L-BFGS with a limit of 1,000 queries per input and number of restarts when it converges. We use sparse GPs to better scale Bayesian optimization \cite{mcintire2016sparse}, as the standard GP scales cubically with the number of observations. We also compare with uniform random parameter selection from Sect.~4. These are untargeted input-specific black-box attacks where the adversary has no knowledge of the target model and only requires the output probability vector of the model.

\textbf{Results.} Table~\ref{table:5_bb_one} shows that Bayesian optimization reached a high 91.6\% success rate with with just 7 queries per successful evasion on average,  under a restrictive 100 query limit. This vastly improves on the query efficiency over random parameters by 5-fold whilst attaining the same accuracy. The improvement for Bayesian optimization increasing the query limit from 100 to 1,000 was very incremental. This suggests that the limited codomain of the procedural noise function is setting an upper bound on the attack's success rate.

L-BFGS performed the worst as we observe it can get trapped and waste queries at \emph{poor} local optima, which requires several restarts from different initial points to improve its performance. There were similar trends for Gabor noise where Bayesian optimization had 78.3\% success with 9.8 queries on average. For both procedural noise functions, Bayesian optimization improved the average query efficiency over L-BFGS and random parameter selection by up to 7 times while retaining a success rate greater than 83\%. We include the other results in Appendix~\ref{appendix:bb}.


\begin{table}[t]
\begin{minipage}{\columnwidth}
	\caption{Comparison on Inception v3 between our input-specific Perlin noise black-box attacks and bandit attack \cite{ilyas2018prior} for different query limits.}
\label{table:5_bb_one}
	\small
\begin{tabular*}{\linewidth}{@{\extracolsep{\fill}} l ccc @{}}
	Attack & Query Limit & Average Queries & Success Rate (\%)\\
	\midrule
	BayesOpt$_{\text{per}}$	& 100 & 7.0 & 91.6\\
	BayesOpt$_{\text{per}}$	& 1,000 & \textbf{8.4} & \textbf{92.8}\\
	L-BFGS$_{\text{per}}$ 	& 100 & 19.8 & 71.7\\
	L-BFGS$_{\text{per}}$	& 1,000 & 70.1& 86.5\\
	Random$_{\text{per}}$	& 1,000 & 36.3\footnote{For each input, we divide the total number of perturbations by the number of those that evade that input to get an expected number of queries.} & 91.6\\
	\midrule
	Bandits$_{\text{TD}}$	& 100 & 29.0 & 36.7\\
	Bandits$_{\text{TD}}$	& 1,000 & 224 & 73.6\\
	Bandits$_{\text{TD}}$	& 10,000 & \textbf{888} & \textbf{96.9}\\
	\bottomrule
\end{tabular*}
\end{minipage}
\end{table}

\textbf{Comparison.} We compare our results with \citet{ilyas2018prior}, where they formalize their attack as a gradient estimation problem (like in white-box gradient-based attacks), and use a bandit optimization framework to solve it. We test the bandits attack on the same model, dataset, and $\ell_{\infty}$-norm for maximum query limits of 100, 1,000, and 10,000.

The results show that the evasion rate of our input-specific procedural noise attack greatly outperforms the bandits attack when the query limit per image is a thousand or less. For significantly larger query limits, the Perlin noise Bayesian optimization attack has a competitive success rate at a drastically better query efficiency--needing \emph{100} times less queries on average.

The procedural noise attack is optimized for universal evasion and a restrictively small amount of queries. Most existing methods explore the entire image space which has a hundred thousand dimensions, and some attacks reduce their search space with techniques like tiling, but the dimensionality is still much larger than the four parameters of our procedural noise. Other input-specific black-box attacks \cite{bhagoji2018black, brendel2017decision, chen2017zoo} require tens to hundreds of thousands of queries on realistic natural-image datasets, which makes them inefficient, but almost certain to find an adversarial example given enough queries. The bandits method makes a small sacrifice in success rate for more query efficiency, and our procedural noise attack takes it further for greater query-efficiency. 

Procedural noise perturbations have naturally high evasion rates, although the expressiveness of our chosen functions can be less than attacks whose larger codomains can capture more kinds of adversarial perturbations. These other attacks make this trade-off by sacrificing efficiency in black-box scenarios, as the number of queries they need to craft successful adversarial examples is large. On the other hand, we can increase the expressiveness of our procedural noise functions by introducing more parameters by, for example, using different colour maps. However, this may come at the cost of its naturally high universal evasion.


\textbf{Summary.} Black-box optimization techniques like Bayesian optimization are able to take advantage of the drastically reduced search space of procedural noise. Together with its naturally high universal evasion, we show that procedural noise gives rise to inexpensive yet potent untargeted black-box attacks, whether it be input-specific or universal. It is also shown to be competitive with existing attacks that often require orders of magnitude more queries or resources.

\section{Other Applications}
In this section, we show how procedural noise attacks extend to object detection DCNs by attacking \emph{YOLO v3} model. We then discuss how exploiting the sensitivity of DCNs to low-level features, like in procedural noise attacks, can be generalized to craft universal attacks for DCNs in other application domains.

\subsection{Attacking Object Detection}
\textit{Object detection} requires the identification of locations and classes of objects within an image. ``Single Shot Detectors'' (SSD) such as the different versions of YOLO \cite{redmon2016you, redmon2017yolo9000} generate their predictions after a single pass over the input image. This allows SSDs to process images in real-time speeds while maintaining high accuracy. Other types of object detectors such as Fast R-CNN \cite{girshick2015fast} and Faster R-CNN \cite{ren2015faster} are based on region proposals. They have comparable accuracy but do not achieve the same image processing speed as SSDs. We test our attack against YOLO v3 \cite{redmon2018yolov3}, the latest iteration of YOLO, which achieves high accuracy and detects objects in real time.

\textbf{Metrics.} We use precision, recall, and mean average precision (mAP) as our primary metrics; mAP is frequently used when comparing the overall performance of object detectors. In object detection the classifier predicts a bounding box to identify the location of an object and assigns that box a class label. The \textit{Intersection Over Union (IOU)} is the ratio of the intersection area over the union area between the predicted and ground truth bounding boxes. A threshold is usually set to determine what IOU constitutes a positive classification. True positives occur when the IOU is larger than the threshold and the class is identified correctly. False negatives occur when the threshold is not met with the correct class for a ground truth bounding box. False positives occur when the IOU is less than the threshold, there are no intersecting ground truth boxes, or there are duplicate predicted bounding boxes.

\textbf{Experiment.} We use the MS COCO dataset \cite{lin2014microsoft} which contains 80 different classes, with a large proportion of targets being the ``person'' class. This has become one of the benchmark datasets for object detection. We use standard settings as described in \cite{redmon2018yolov3}, with input dimensions $416 \times 416 \times 3$ and an IOU threshold of 0.5.

We use $\ell_{\infty}$-norm $\varepsilon = 16$ and apply each of our procedural noise perturbations on the 1,000 random images from the validation set. The perturbations generated are from 1,000 Gabor noise, 1,000 Perlin noise, and 1,000 $\ell_{\infty}$-optimized uniform random perturbations. The parameters are chosen uniformly at random so that we can analyze the results, as in Sect.~4. This is a universal untargeted black-box attack.

\begin{figure}[htb]
	\centering
	\includegraphics[clip, width = 3.35in]{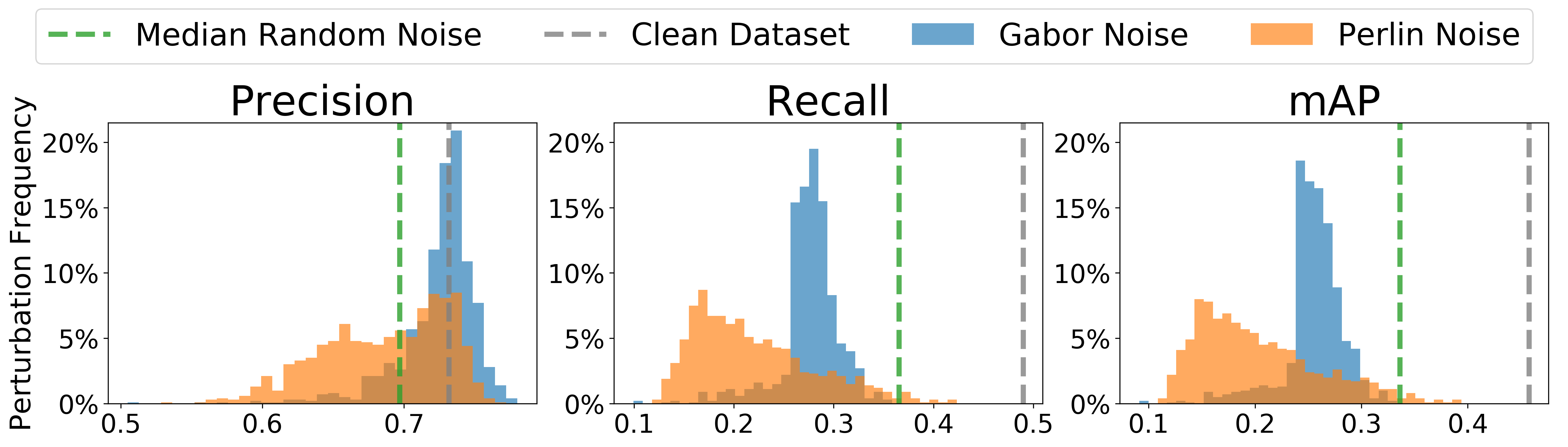}
	\includegraphics[clip, width = 3.35in]{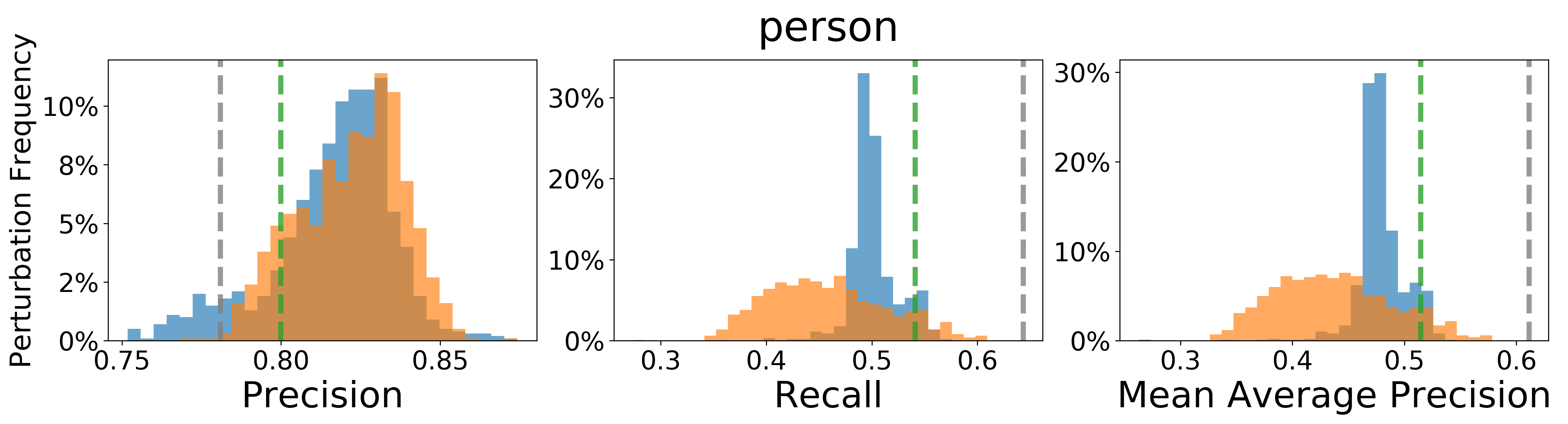}
	\caption{Histogram of metrics over all perturbations on YOLO v3. Results for (top) all classes and (bottom) the ``person'' class.}
	\label{fig:6_yolo}
\end{figure}

\textbf{Results.} The procedural noise perturbations significantly decrease the mAP of the model, with 75\% of Perlin noise perturbations halving the mAP or worse. The precision and recall also decreased in the presence of our perturbations. Again, Perlin noise outperforms Gabor noise on average but Gabor noise has a smaller minimum, with at least one perturbation causing 0.09 mAP. Compared to uniform random noise, which maintained around 0.34 mAP, both Perlin and Gabor noise had larger impact on the model's performance. We represent uniform random noise using the median (which, in this case, is very close to the mean) on each metric as the variance is very small (less than $10^{-5}$).

In Fig.~\ref{fig:6_yolo}, for the metrics across all classes, we can observe that the precision was maintained or decreased while the recall and mAP decreased. Some perturbations decreased all three metrics, which indicate that these cause both false positives and false negatives consistently. When the perturbation increases the precision while decreasing the recall and mAP, then the noise is likely masking objects rather than introducing new objects to the image. This masking effect has serious implications for security applications like surveillance and autonomous vehicles.

We focus on the ``person'' class as it constitutes a majority of the targets and is semantically meaningful in the context of some applications. In Fig.~\ref{fig:6_yolo}, metrics for ``person'' follow a similar trend to the metrics across all classes. However, the increase in precision is very small (<0.10) compared to the large drops in recall and mAP caused by Perlin noise. The higher precision indicates fewer false positives, while the decrease in recall indicates more false negatives. This indicates that the classifier is making fewer predictions overall, which means that the noise is masking persons in the image.

Whilst for the most relevant ``person'' class, our procedural noise appears to have an obfuscating effect, for other classes like ``zebra'' all the metrics decrease -- indicating that there are many false positive and false negatives. However, for three classes, ``backpack'', ``book'', and ``toaster'', all three metrics improve. These labels did not have as much representation in the test set which may explain this anomalous result.

The frequency of the sine $\phi_{\text{sine}}$ for Perlin noise had the largest inverse correlation of less than -0.72 with each of the three metrics. This means that high-frequency patterns decrease the model's performance metrics, similarly to what we have observed for the image classifiers. The thickness $\lambda$ of Gabor noise is moderately correlated at 0.4 with the precision, but not the recall or mAP. This suggests that thicker Gabor noise perturbations decrease the number of false positives relative to the other perturbations.

\textbf{Discussion.} Object detection is a more complex task than image classification as it has to identify multiple objects and their locations within an image. Although YOLO v3 has a different architecture, task, and dataset from the ImageNet classifiers, we see that the same procedural noise perturbations are still able to greatly degrade its performance. This shows that it is more likely caused by the models' sensitivity towards perturbations rather than a texture bias, as the object detection task has a spatial component to it. This suggests that our procedural noise attacks may generalize to other DCNs on computer vision tasks with natural-image data.

In our discussion in Sect.~4.4, we hypothesized that the procedural noise perturbations are an aggregation of low-level features at high frequencies that the DCNs strongly respond to. The prior that convolutional layers induce and similarities across natural-image datasets may be why DCNs are sensitive to these noise patterns. Additionally, like in Sect.~5, we can also create more efficient black-box attacks by applying black-box optimization techniques such as Bayesian optimization to enhance the attack against object detectors. When using Bayesian optimization, an attacker can focus on minimizing a specific metric (precision, recall, mAP, or F1 score).

\subsection{Beyond Images}
One of the main strengths of procedural noise is that it allows to describe a distribution of UAPs with only a few parameters. This compressed representation allows for an efficient generation of UAPs both for undermining a machine learning service or, defensively, for testing the robustness of a model. In practice, compressed representations can be learned by generative models such as GANs \cite{mopuri2018nag, xiao2018generating} or Variational Autoencoders (VAEs), however these incur additional training, calibration, and maintenance costs. Training algorithms or defences that incorporate domain-specific knowledge may be needed to mitigate the sensitivity towards attacks that make use of these compact representations.


DCNs in other applications may also be vulnerable to aggregations of low-level features due to the use of convolutional layers. Future attacks can exploit how DCNs rely on combining low-level features rather than understanding the more difficult global features in the input data. A natural next step would be to apply these ideas in exploratory attacks or sensitivity analysis on sensory applications like speech recognition and natural language processing. To expand our procedural noise attack framework to other applications, it is worth identifying patterns in existing adversarial examples for domains like speech recognition \cite{carlini2016hidden, carlini2018audio} and reinforcement learning \cite{huang2017adversarial, lin2017tactics} to find analogues of procedural noise. As a starting point, these patterns can be found by finding perturbations that maximize the hidden layer difference as in the Singular Vector Attack \cite{khrulkov2018art} or by using feature visualization on earlier layers of DCNs to infer the low-level features that a model learns.



\section{Preliminary Defence}
The DCNs we tested are surprisingly fragile to procedural noise as UAPs. Improving their robustness to adversarial examples is not a straightforward task. A robust defence needs to defend not only against existing attacks but also future attacks, and often proposed defences are shown to fail against new or existing attacks \cite{carlini2019evaluating}.

Among existing defences, adversarial training appears to be more robust than others \cite{athalye2018obfuscated}. However, we have shown in Table~\ref{table:4_input-spec} that ensemble adversarial training against gradient-based attacks did not significantly diminish the input-specific evasion rate of our procedural noise attack. This suggests that such defences do not generalize well as the attacks used for adversarial training do not sufficiently represent the entire space of adversarial examples. Training against all types of adversarial attacks would become computationally expensive, especially for high-dimensional tasks like ImageNet. Thus, defences that regularize the model or incorporate domain-specific knowledge may be more efficient strategies.

DCNs' weakness to adversarial perturbations across inputs may be a result of their sensitivity---small changes to the input cause large changes to the output. Thus, input-agnostic defences that minimize the impact of small perturbations may be effective in reducing the models' sensitivity without the need to train against all types of attacks. As a preliminary investigation, we briefly explore here using denoising to defend against the universality of our procedural noise perturbations.

\begin{figure}[H]
	\centering
	\includegraphics[clip, width = 3.35in]{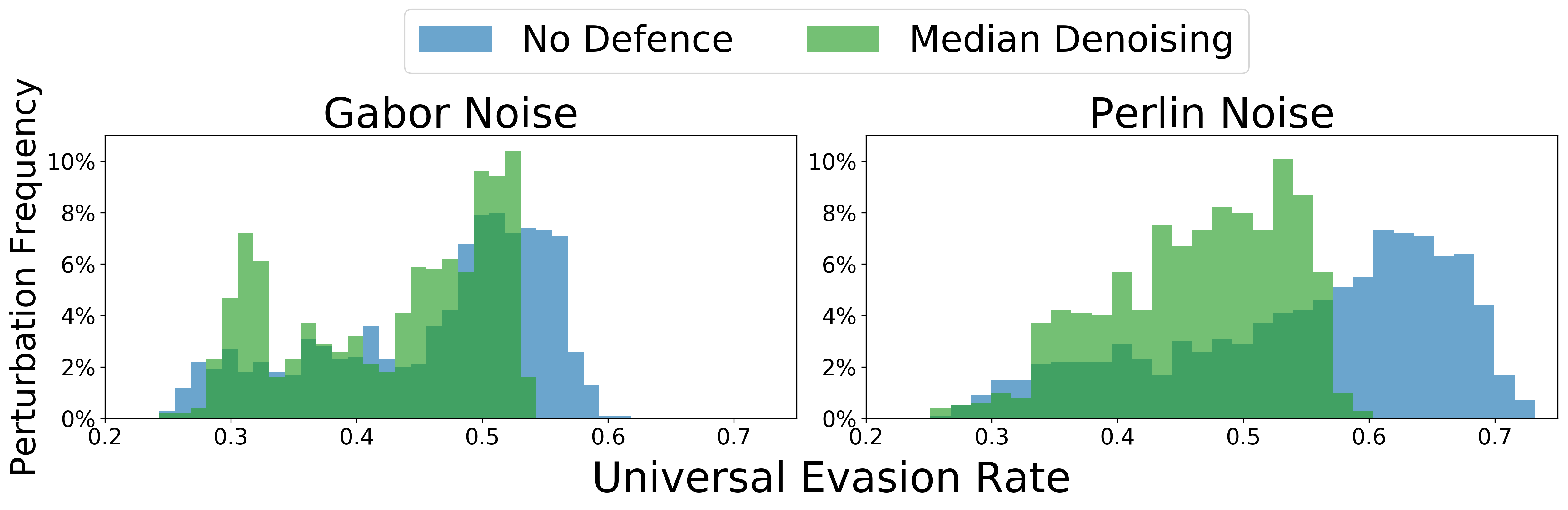}
	\caption{Histogram of metrics over all perturbations on Inception v3 with and without median denoising for (left) Gabor noise and (right) Perlin noise for $\ell_{\infty}$-norm $\varepsilon = 16$.}
	\label{fig:7_defences}
\end{figure}

\subsection{Denoising}
Denoising with spatial filters is a common pre-processing step in signal processing and is thus an attractive defence due to its simplicity and pervasiveness across signal processing applications. Median filtering is a smoothing operation that replaces each entry with the median of its neighbouring entries, and it often preserves edges while removing noise. The idea is to smooth out the high-frequency noise, which we know to be highly correlated with the universal evasion rates for Perlin noise on Inception v3.

\textbf{Experiment.} We retrain Inception v3 with a median denoising filter applied as a pre-processing step. For this evaluation, we use the same 1,000 of each procedural noise, 1,000 uniform random perturbations, and 5,000 validation set points as in Sect.~4. We test how denoising improves robustness for $\ell_{\infty}$ attack norms $\varepsilon = 4, 8, 12,$ and 16. The decrease in the number of uniform random perturbations is due to results in Sect.~4, where we found that its universal evasion rate had very low variance across 10,000 samples, and our results show it is similar for 1,000 samples.

\textbf{Results.} The min-max oscillations present in the procedural noise perturbations may have allowed the noise patterns to persist despite the denoising, as the universal evasion rates are not completely mitigated by the defence. Across the different $\ell_{\infty}$-norm values, the denoising decreased the median and mean universal evasion rates by a consistent amount when compared with no defence: 7.2-10.8\% for Gabor noise and 13.2-16.9\% for Perlin noise. Fig.~\ref{fig:7_defences} shows that the decrease in effectiveness of Perlin noise is much greater than that for Gabor noise, suggesting that the denoising appears to slightly mitigate the effectiveness of high-frequency noise patterns.

This denoising defence grants a reasonable increase in robustness despite the simplicity of the algorithm. However, this measure has not fully mitigated the sensitivity to procedural noise. We hope future work explore more input-agnostic defences that minimize the sensitivity of large-scale models to small perturbations with techniques like model compression \cite{han2015deep}, Jacobian regularization \cite{rifai2011contractive, sokolic2017robust, jakubovitz2018improving, khrulkov2018art}, or other types of denoising.

\section{Conclusion}
We highlight the strengths of procedural noise as an indiscriminate and inexpensive black-box attack on DCNs. We have shown that popular DCN architectures for image classification have systemic vulnerabilities to procedural noise perturbations, with Perlin noise UAPs that achieve 58\% or larger universal evasion across non-adversarially trained models. This weakness to procedural noise can be used to craft efficient universal and input-specific black-box attacks. Our procedural noise attack augmented with Bayesian optimization attains competitive input-specific success rates whilst improving the query efficiency of existing black-box methods over 100 times. Moreover, we show that procedural noise attacks also work against the YOLO v3 model for object detection, where it has an obfuscating effect on the ``person'' class.

Our results have notable implications. The universality of our black-box method introduces the possibility for large-scale attacks on DCN-based machine learning services, and the use of Bayesian optimization makes untargeted black-box attacks significantly more efficient. We hypothesize that our procedural noise attacks exploit low-level features that DCNs are sensitive towards. If true, this has worrying implications on the safety of transfer learning, as it is often these low-level features that are preserved when retraining models for new tasks. It may be the case that universal attacks can be extended to other application domains by using compact representations of UAPs that exploit the sensitivity of models to low-level features.

We have shown the difficulty of defending against such novel approaches. In particular, ensemble adversarial training on gradient-based methods was unable to significantly diminish our procedural noise attack. We suggest that future defences take more input-agnostic approaches to avoid the costs in defending and retraining against all possible attacks. Our work prompts the need for further research of more intuitive and novel attack frameworks that use analogues of procedural noise in other machine learning application domains such as audio processing and reinforcement learning. We hope future work will explore in more depth the nature of cross-model universal adversarial perturbations, as these vulnerabilities generalize across both inputs and models, and a more formal framework could better explain why our procedural noise attacks are effective.


\bibliographystyle{ACM-Reference-Format}
\bibliography{proc_noise}


\appendix

\section{Examples of Perturbations}
\label{appendix:ex_perturbs}
\begin{figure}[H]
	\centering
	\includegraphics[clip, width = 3.35in]{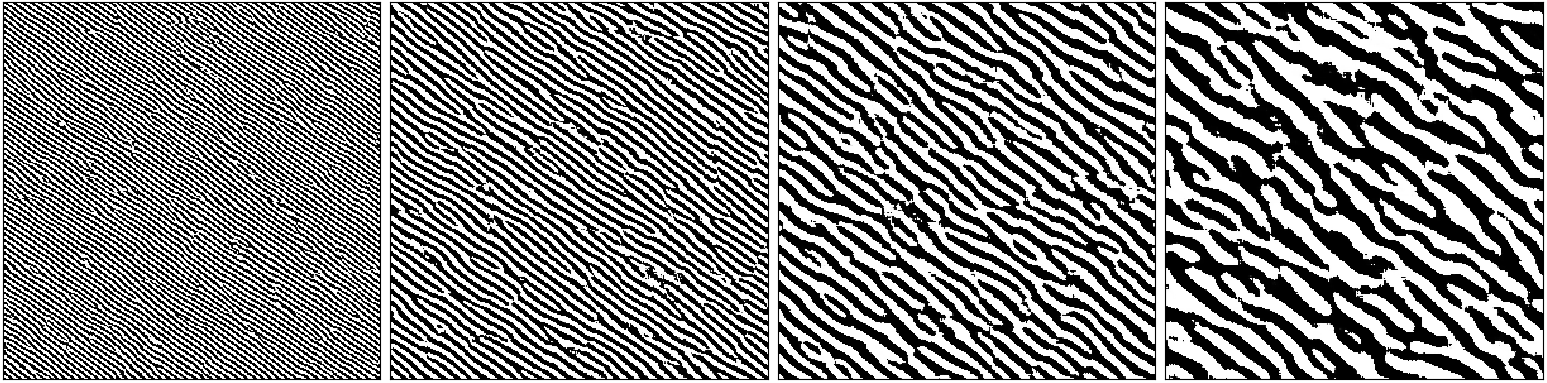}
	\includegraphics[clip, width = 3.35in]{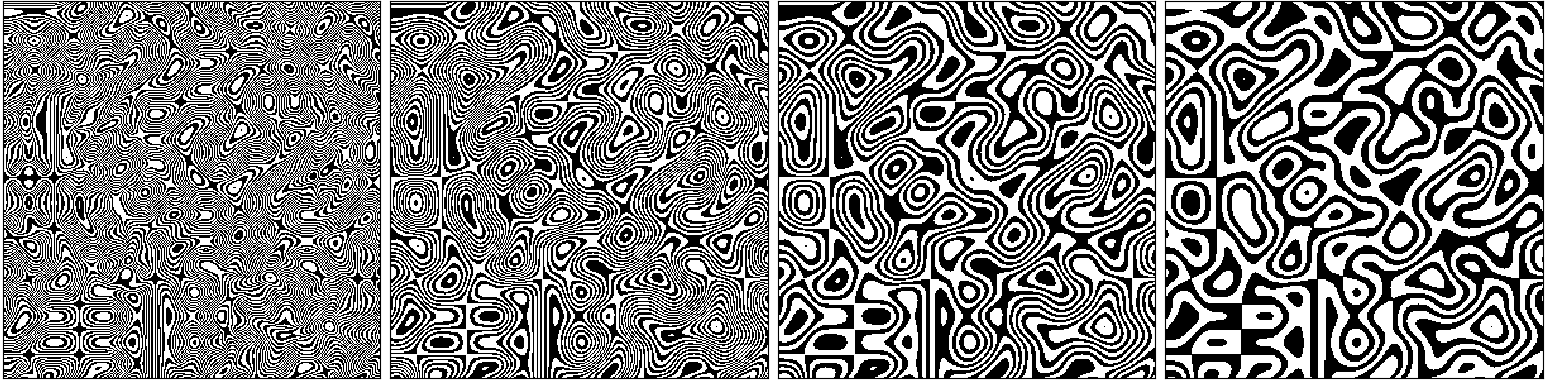}
	\caption{Procedural noise patterns with (top) Gabor noise and (bottom) Perlin noise, both with decreasing frequency from left to right.}
	\label{fig:3_gabor}
\end{figure}

\begin{figure}[H]
	\centering
	\includegraphics[clip, width = 3.35in]{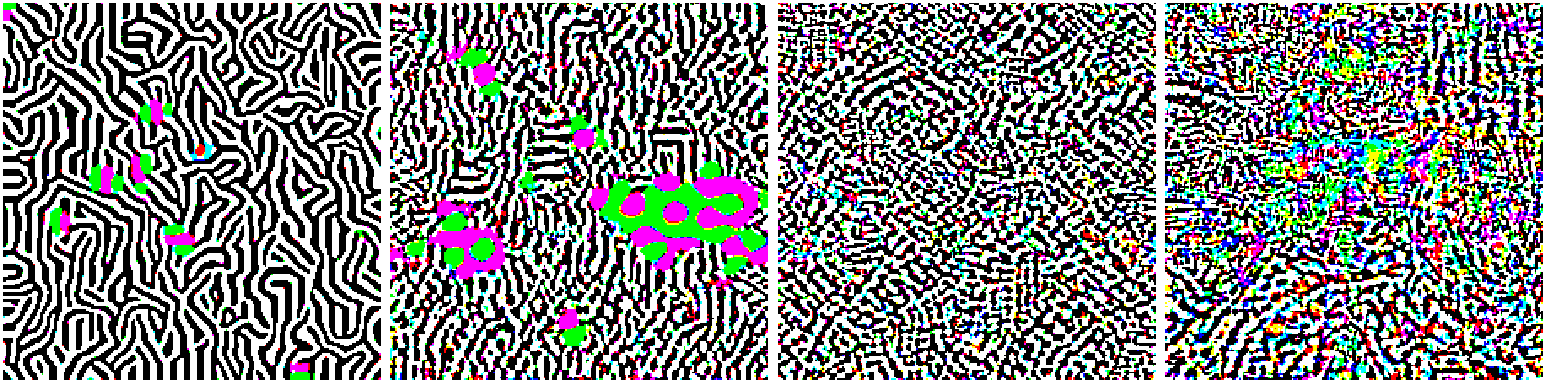}
	\caption{UAPs generated for VGG-19 using the white-box Singular Vector Attack \cite{khrulkov2018art}.}
	\label{fig:3_uap_khrulkov}
\end{figure}

Figs.~\ref{fig:3_gabor} and \ref{fig:3_uap_khrulkov} show examples of perturbations from procedural noise functions and existing white-box UAP attacks. Notice the visual similarities in structure between the noise patterns in Figs.~\ref{fig:3_gabor} and \ref{fig:3_uap_khrulkov}.

\section{Top 5 Input-Specific Evasion}
\label{appendix:top5}
In Table~\ref{table:4_input-spec5}, we consider the top 5 input-specific evasion rate where the true class label is outside the top 5 class labels in the perturbed image $x + s$. The top 5 error is often used alongside the top 1 error to measure the performance of classifiers on ImageNet. Top 5 evasion is more difficult than top 1 as the confidence in the true label has to be degraded sufficiently enough for it to be below five other class labels. Despite that, Perlin noise is able to achieve top 5 evasion on more than half the inputs for all tested models. We see that our procedural noise still remains a strong attack for the top 5 error metric.

\begin{table}[H]
	\caption{Top 5 input-specific evasion rate (in \%) for random and procedural noise perturbations. Original refers to the top 5 error on the unaltered original images. The strongest attack on each classifier is highlighted.}
	\label{table:4_input-spec5}
	\small
	\begin{tabular*}{3in}{@{\extracolsep{\fill}} l cccc @{}}
		Classifier & Original & Random & Gabor & Perlin\\
		\midrule
		VGG-19			& 9.5 & 28.0 & 90.2 & \textbf{92.6}\\
		ResNet-50		& 8.4 & 29.5 & 82.5 & \textbf{85.3}\\
		Inception v3		& 6.3 & 18.0 & 66.9 & \textbf{79.5}\\
		IRv2				& 4.5 & 13.2 & 56.3 & \textbf{66.4}\\
		IRv2$_{\text{ens}}$	& 5.1 & 12.1 & 44.1 & \textbf{53.6}\\
		\bottomrule
	\end{tabular*}
\end{table}

\section{Correlation Matrices}
\label{appendix:corr_mat}
\begin{figure}[H]
	\centering
	\includegraphics[width = 2.6in]{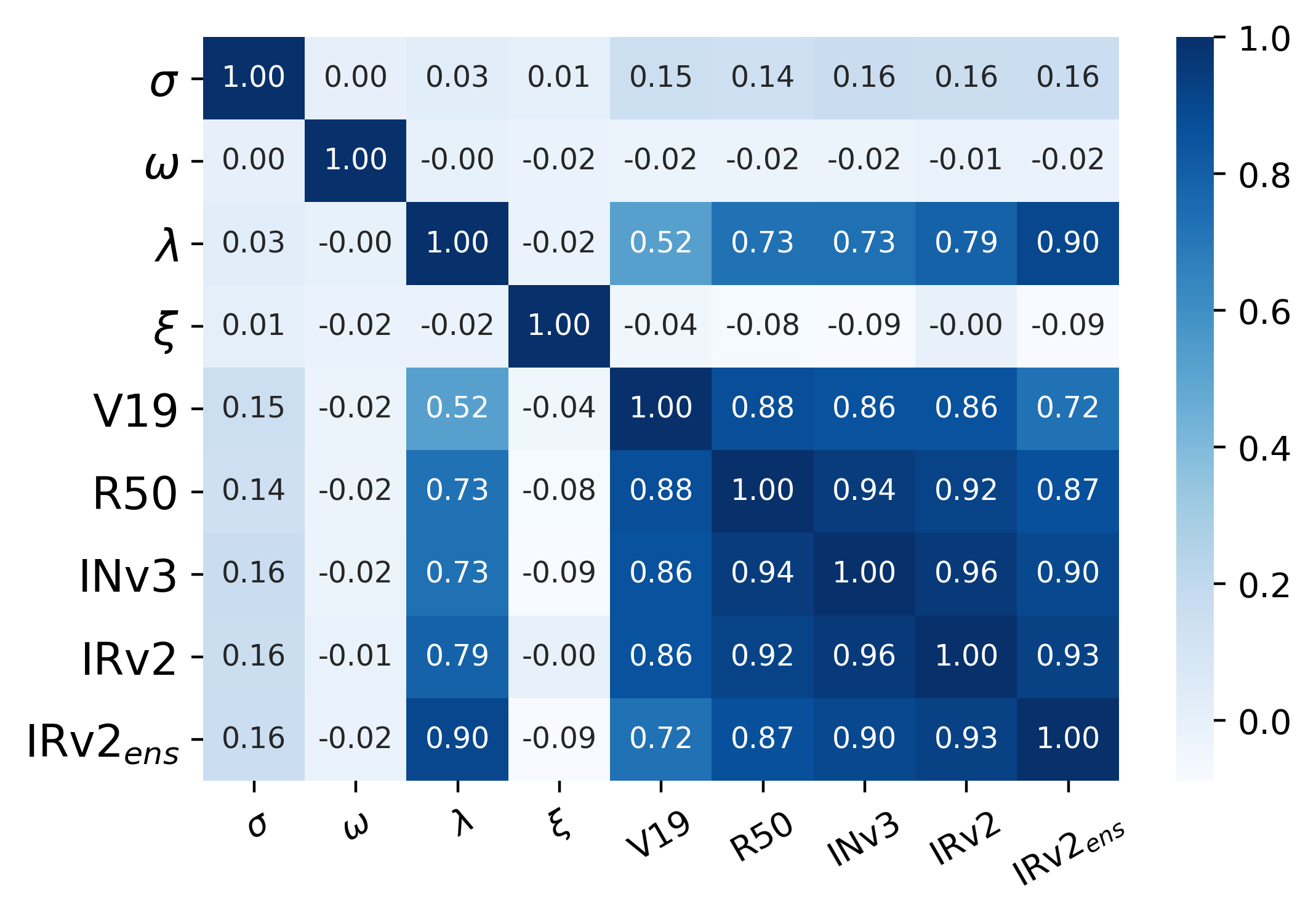}
	\caption{Correlations between Gabor noise parameters ($\delta_{\text{gab}} = \{\sigma, \omega, \lambda, \xi\}$) and universal evasion rates on each image classifier.}
	\label{fig:4_corr_mat_gba}
\end{figure}

\begin{figure}[H]
	\centering
	\includegraphics[width = 2.6in]{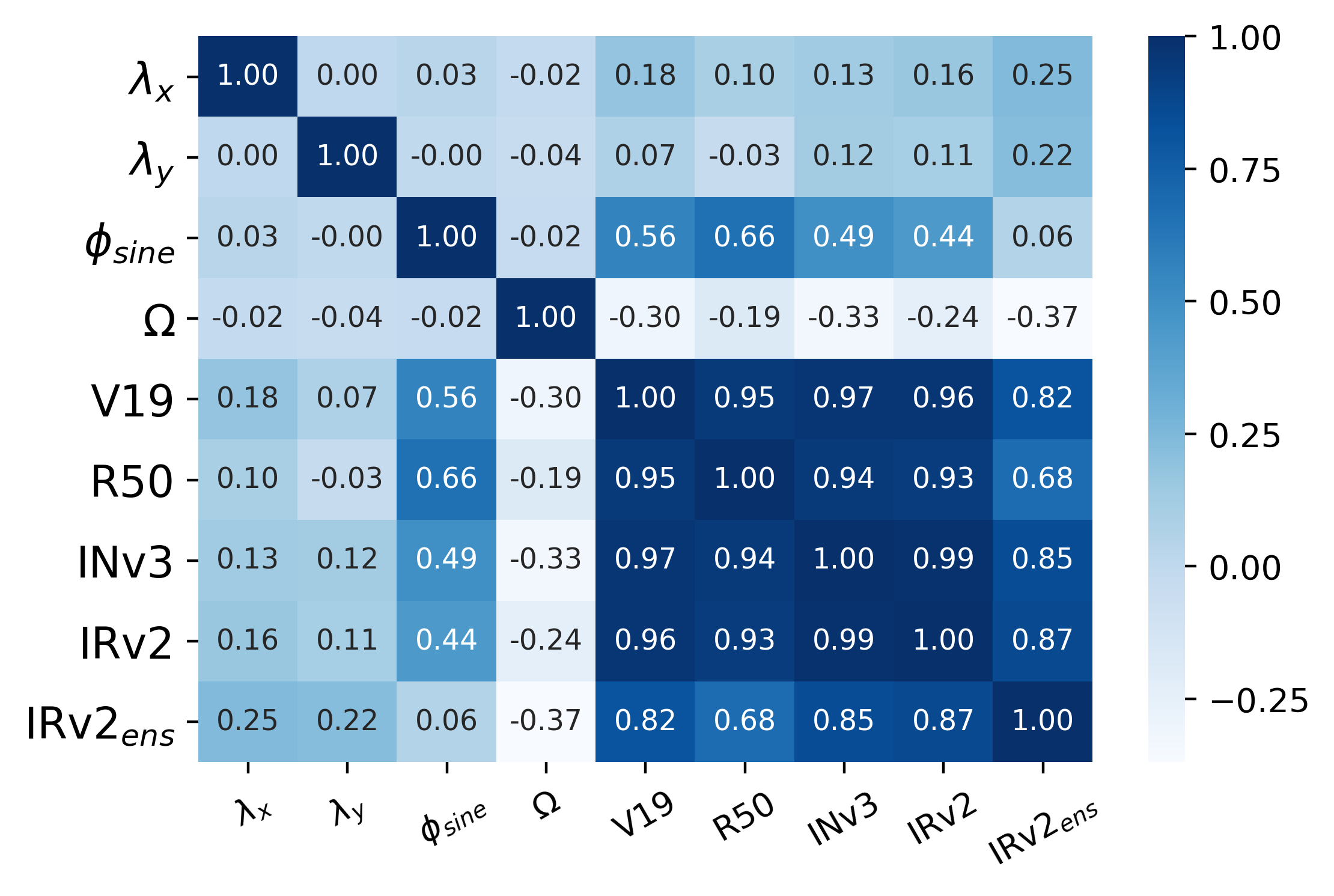}
	\caption{Correlations between Perlin noise parameters ($\delta_{\text{per}} = \{\lambda_x, \lambda_y, \phi_{\text{sine}}, \omega\}$) and universal evasion rates on each image classifier.}
	\label{fig:4_corr_mat_per}
\end{figure}

Procedural noise perturbations had their parameters chosen uniformly at random and with constraint $\ell_{\infty}$-norm $\varepsilon = 16$. For Sect.~4, Figs.~\ref{fig:4_corr_mat_gba} and \ref{fig:4_corr_mat_per} show the correlation matrices between the procedural noise parameters and the corresponding universal evasion rates on the various ImageNet classifiers. These correlations quantify the effects of each parameter and the transferability of UAPs across models. The labels ``V19'', ``R50'', and ``INv3'' refer to VGG-19, ResNet-50, and Inception v3 respectively.


%


\section{Black-box Attack Results}
\label{appendix:bb}
For Sect.~5, full results for input-specific and universal black-box attacks against Inception v3 are in Tables~\ref{table:bb_one_full} and \ref{table:bb_uap_full}. We include results for Gabor noise (gab) and Perlin noise (per). The bandit attacks \cite{ilyas2018prior} include the use of time (T) and time with data (TD) priors. Methods are separated according to the base attack that was used: Gabor noise, Perlin noise, and bandits framework.

\begin{table}[H]
\begin{minipage}{\columnwidth}
	\caption{Results for input-specific black-box attacks on Inception v3 with $\ell_{\infty}$-norm $\varepsilon = 16$. Gabor and Perlin noise attacks are labeled with ``gab'' and ``per'' respectively. Best success rate per method is highlighted.}
	\label{table:bb_one_full}
	\small
\begin{tabular*}{\linewidth}{@{\extracolsep{\fill}} l ccc @{}}
	Attack & Query Limit & Average Queries & Success Rate (\%)\\
	\midrule
	BayesOpt$_{\text{gab}}$	& 100 & 9.8 & 83.1\\
	BayesOpt$_{\text{gab}}$	& 1,000 & 10.9 & 83.6\\
	L-BFGS$_{\text{gab}}$ 	& 100 & 6.0 & 44.7\\
	L-BFGS$_{\text{gab}}$	& 1,000 & 6.0 & 44.7\\
	Random$_{\text{gab}}$	& 1,000 & \textbf{62.2} & \textbf{86.1}\\
	\midrule
	BayesOpt$_{\text{per}}$	& 100 & 7.0 & 91.6\\
	BayesOpt$_{\text{per}}$	& 1,000 & \textbf{8.4} & \textbf{92.8}\\
	L-BFGS$_{\text{per}}$ 	& 100 & 19.8 &  71.7\\
	L-BFGS$_{\text{per}}$	& 1,000 & 70.1 &  86.5\\
	Random$_{\text{per}}$	& 1,000 & 36.3 & 91.6\\
	\midrule
	Bandits$_{\text{T}}$		& 100 & 31.3 & 14.3\\
	Bandits$_{\text{T}}$		& 1,000 & 330 & 47.6\\
	Bandits$_{\text{T}}$		& 10,000 & 1,795 & 88.3\\
	Bandits$_{\text{TD}}$	& 100 & 29.0 & 36.7\\
	Bandits$_{\text{TD}}$	& 1,000 & 224 & 73.6 \\
	Bandits$_{\text{TD}}$	& 10,000 & \textbf{888} & \textbf{96.9} \\
	\bottomrule
\end{tabular*}
\end{minipage}
\end{table}

\begin{table*}[htb]
	\caption{Results for universal black-box attacks on Inception v3 with $\ell_{\infty}$-norm $\varepsilon = 16$. Gabor and Perlin noise attacks are labeled with ``gab'' and ``per'' respectively. Universal evasion rates (\%) of the optimized perturbations are shown for their respective training set and the validation set.}
\label{table:bb_uap_full}
	\small
\begin{tabular*}{5.6in}{@{\extracolsep{\fill}} l cccccccccccccc @{}}
	\toprule
	& \multicolumn{2}{c}{BayesOpt$_{\text{gab}}$} && \multicolumn{2}{c}{L-BFGS$_{\text{gab}}$} && \multicolumn{2}{c}{BayesOpt$_{\text{per}}$} && \multicolumn{2}{c}{L-BFGS$_{\text{per}}$}\\
	\cmidrule{2-3} \cmidrule{5-6} \cmidrule{8-9}  \cmidrule{11-12}
	Train Size & Train & Val. && Train & Val. && Train & Val. && Train & Val.\\
	\midrule
	50		& 64.0 & 57.6 && 58.0 & 51.6 && 78.0 & 71.4 && 74.0 & 69.9\\
	125		& 64.6 & 58.0 && 58.4 & 54.6 && 77.6 & 70.2 && 76.0 & 71.5\\
	250		& 60.0 & 58.4 && 58.8 & 56.0 && 71.6 & 71.2 && 71.2 & 69.7\\
	500		& 64.8 & 62.4 && 59.8 & 58.0 && 75.0 & 72.9 && 73.4 & 70.8\\
	\bottomrule
\end{tabular*}
\end{table*}

\newpage

\section{Adversarial Examples}
\label{appendix:ex_adv}
\begin{figure}[htb]
	\centering
	\includegraphics[clip, width = 3.35in]{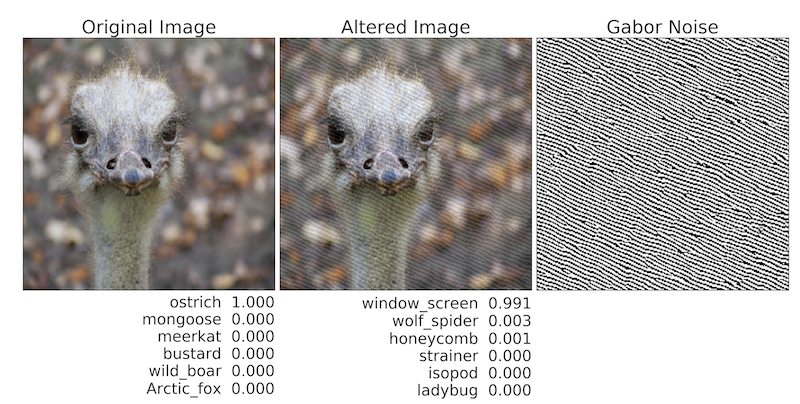}
	\includegraphics[clip, width = 3.35in]{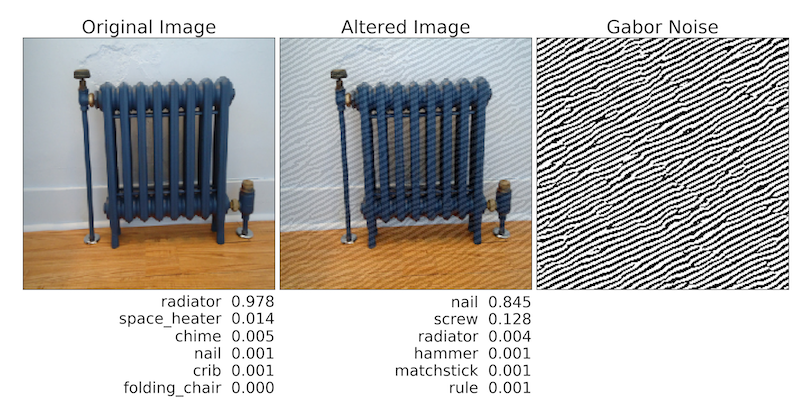}
	\includegraphics[clip, width = 3.35in]{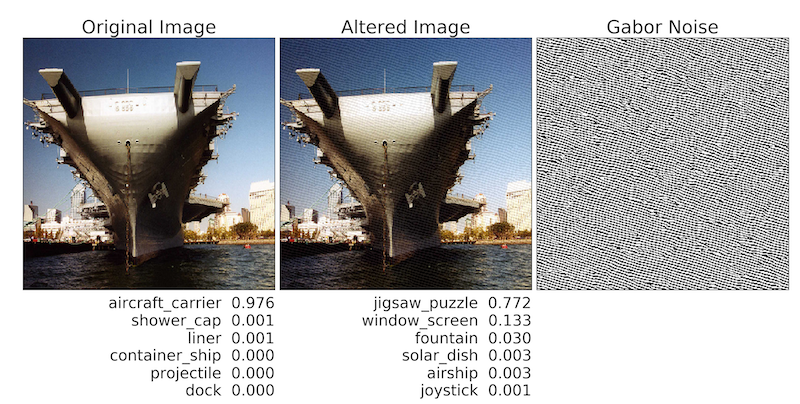}
	\caption{Gabor noise adversarial examples on (top) Inception v3, (middle) VGG-19, and (bottom) ResNet-50.}
	\label{fig:ex_img}
\end{figure}

\begin{figure}[htb]
	\centering
	\includegraphics[clip, width = 1.65in]{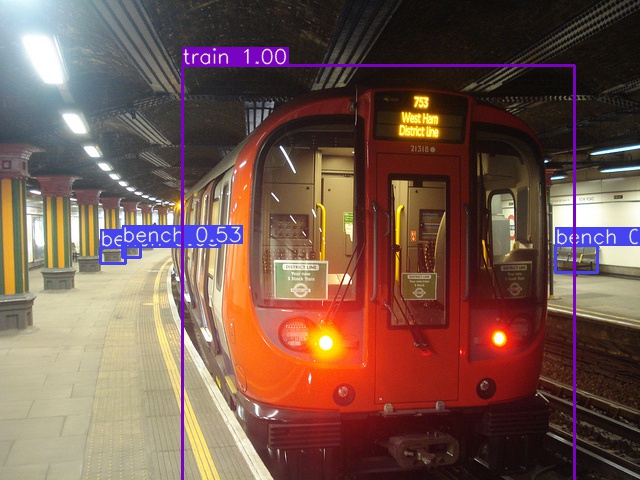} \includegraphics[clip, width = 1.65in]{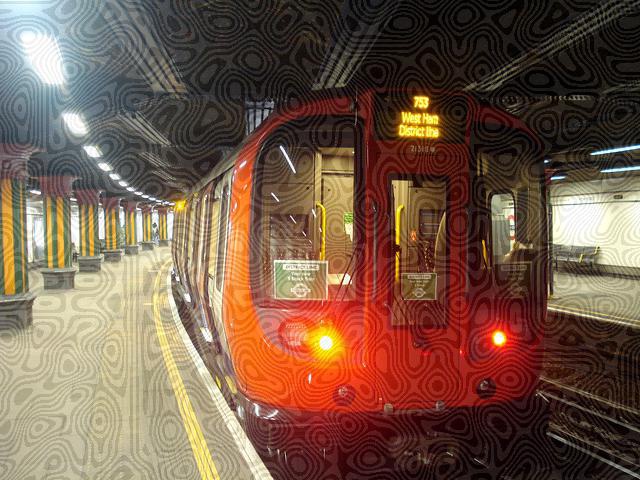}
	\includegraphics[clip, width = 1.65in]{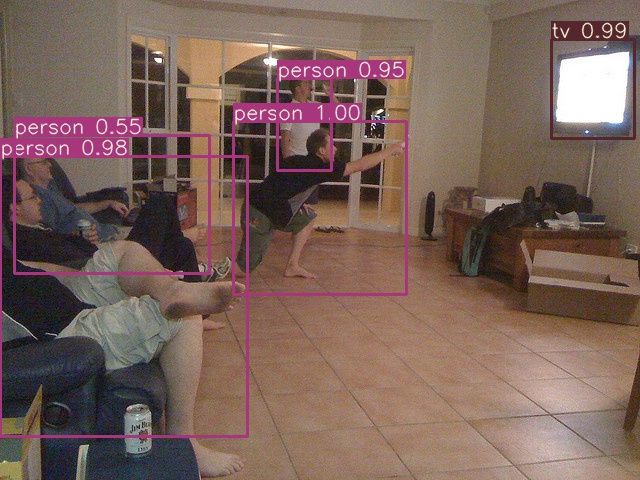} \includegraphics[clip, width = 1.65in]{yolo_perlin/6}
	\includegraphics[clip, width = 1.65in]{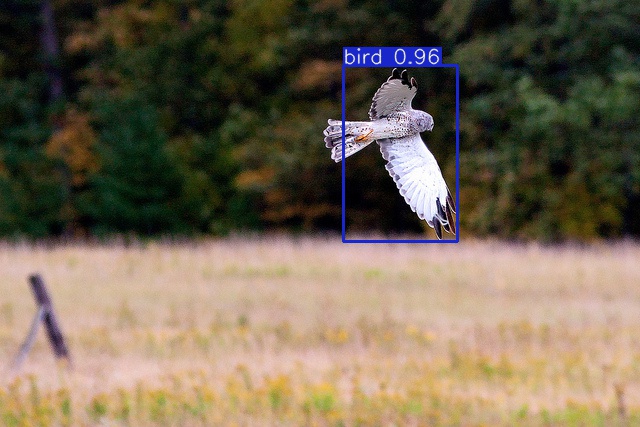} \includegraphics[clip, width = 1.65in]{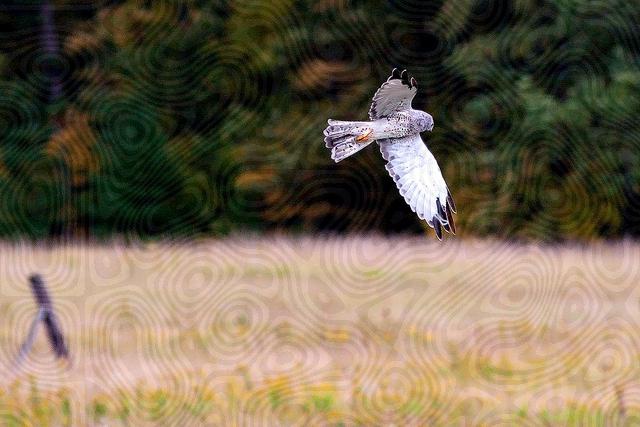}
	\includegraphics[clip, width = 1.65in]{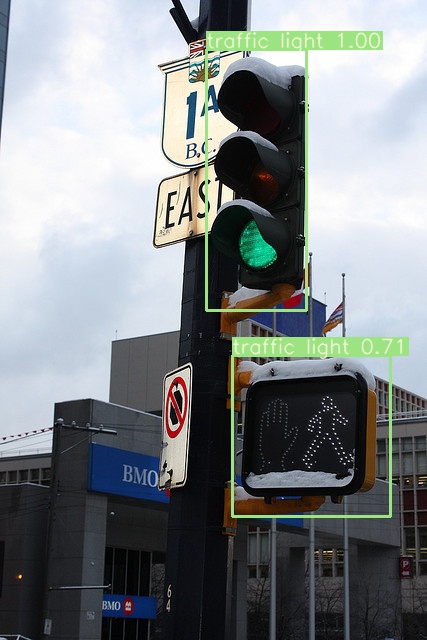} \includegraphics[clip, width = 1.65in]{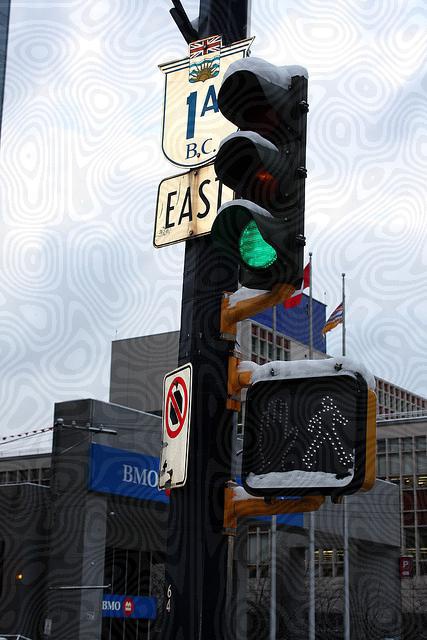}
	\caption{Perlin noise adversarial examples on YOLO v3.}
	\label{fig:ex_yolo}
\end{figure}

Figs.~\ref{fig:ex_img} and \ref{fig:ex_yolo} contain some samples of adversarial examples from our procedural noise attacks against image classification and object detection tasks.

\end{document}